\documentclass{article}[11pt]
\usepackage{amsmath,amssymb,epsfig}
\newcommand{\ep}{\epsilon}
\renewcommand{\P}{\phi}
\newcommand{\phm}{\phantom{-}}

\newcommand{\T}{\mathrm{Tr}}

\newcommand{\q}{\theta}
\newcommand{\bq}{\bar \theta}
\renewcommand{\a}{\alpha}
\renewcommand{\b}{\beta}

\begin{document}

\thispagestyle{empty}

\begin{flushright}
LU-ITP 2009/001
\end{flushright}

\vskip 3.0 cm

\begin{center}

{\Large \textbf{Operator mixing in ${\cal N} = 4$ SYM:
\vskip 0.2 cm
The Konishi anomaly re-re-visited}}

\vskip 2.0 cm

\textbf{B.~Eden}

\vskip 0.5 cm

{\it Institut f\"ur theoretische Physik, Universit\"at Leipzig \\
Postfach 100920, D-04009 Leipzig, Germany \\
burkhard.eden@itp.uni-leipzig.de}

\vskip 3 cm

\textbf{Abstract:}
\end{center}
\vskip -0.2 cm
\hskip \parindent
The supersymmetry transformation relating the Konishi operator to its lowest descendant in the
10 of $SU(4)$ is not manifest in the ${\cal N}=1$ formulation of the theory but rather uses
an equation of motion. On the classical level one finds one operator, the unintegrated chiral
superpotential. In the quantum theory this term receives an admixture by a second operator,
the Yang-Mills part of the Lagrangian. It has long been debated whether this ``anomalous''
contribution is affected by higher loop corrections. We present a first principles calculation
at the second non-trivial order in perturbation theory using supersymmetric dimensional
reduction as a regulator and renormalisation by Z-factors. Singular higher loop corrections
to the renormalisation factor of the Yang-Mills term are required if the conformal properties
of two-point functions are to be met. These singularities take the form determined in
preceding work on rather general grounds. Moreover, we also find non-vanishing finite terms.

The core part of the problem is the evaluation of a four-loop two-point correlator which
is accomplished by the Laporta algorithm. Apart from several examples of the T$_1$
topology with two lines of non-integer dimension we need the first few orders in the
$\epsilon$ expansion of three master integrals. The approach is self-contained
in that all the necessary information can be derived from the power counting
finiteness of some integrals.

\newpage

\section{Introduction}

In the ${\cal N} = 4$ super Yang-Mills theory (${\cal N} = 4$
SYM), half BPS operators are known to be protected.
Their two- and three-point functions receive no quantum
corrections other than contact terms \cite{usprotect,contact}.
The simplest such operator is
\begin{equation}
{\cal O} \, = \, \mathrm{Tr}(\phi^1 \phi^1)
\end{equation}
where the notation refers to the action of the theory
(\ref{action}) in terms of ${\cal N}=1$ superfields \cite{bible}.
The physical fields are the $\theta = \bar \theta = 0$ components
of the three complex chiral fields and of
\begin{eqnarray}
\lambda^1_\alpha = & \lambda_\alpha & = \, \frac{1}{4 \, g} \, \bar D_{\dot \alpha}
\bar D^{\dot \alpha} \left(e^{g V} D_\alpha e^{- g V} \right) \\
\lambda^i_\alpha = & \nabla_\alpha \phi^{i-1} & = \, e^{g V} \left( D_\alpha \left( e^{-g V}
\phi^{i-1} e^{g V} \right) \right) e^{- g V} \, , \qquad i \, \in \, \{2,3,4\}
\nonumber
\end{eqnarray}
and their complex conjugates, while the gauge field is in the
$[D_\alpha, \bar D_\alpha]$ component of $V$. The little $i \in
\{1 \ldots 4 \}$ is an index in the fundamental representation of
the non-manifest but unbroken $SU(4)$ R-symmetry of the theory.
Using the equation of motion
\begin{equation}
- \frac{1}{4} \, D^\alpha D_\alpha \, \phi^I \, = \, \frac{g}{2}
\, \varepsilon^{IJK} \, [ \bar \phi_J, \bar \phi_K ] \, , \qquad I
\, \in \, \{1,2,3\} \label{chireom}
\end{equation}
of the chiral field we find
\begin{eqnarray}
\frac{1}{2} \, D^\alpha D_\alpha \, {\cal O} & = & \check F - 4 \, g \, \check B \, ,
\label{Fcheck} \\ \check F & = & \mathrm{Tr} \left( \lambda^{2 \alpha} \lambda^2_\alpha
 \right) \, , \qquad \check B \,
= \, \mathrm{Tr} \left( \left( e^{- g V} \phi^1 e^{g V} \right) \,
\left[ \bar \phi_2, \, \bar \phi_3 \right] \right) \, . \nonumber
\end{eqnarray}
The operator $\check F$ is an $SU(4)$ component of
\begin{equation}
F^{10, ij} \, = \, \mathrm{Tr} \left( \lambda^{i \alpha}
\lambda^j_\alpha \right) \, ,
\end{equation}
and similarly for $\check B$. The associated highest weight components are
\begin{equation}
F \, = \, \mathrm{Tr} \left( \lambda^{\alpha} \lambda_\alpha \right) \, , \qquad
B \, = \, \mathrm{Tr} \left( \phi^1 [ \phi^2, \phi^3 ] \right) \, .
\end{equation}
We verify by explicit graph calculations at $O(g^2)$ in Section 3
for the highest weight components, and at $O(g^4)$ in Section 5
for the $22$ components $\check F, \, \check B$, that the
combination
\begin{equation}
O^{10} \, = \, F^{10} - 4 \, g \, B^{10}
\end{equation}
is a protected operator. Further, the contact contributions in its
two-point function are related to those of the ${\cal O}$ case by
the classical e.o.m. (\ref{chireom}), see Section 5.

Second, from the complex conjugate of the classical relation (\ref{chireom})
we find
\begin{equation}
- \frac{1}{4} \, \bar D_{\dot \alpha} \bar D^{\dot \alpha} \,
{\cal K} \, = \, - 3 \, g \, B \, , \qquad  {\cal K} \, = \,
\mathrm{Tr} \left( e^{g V} \bar \phi_I \, e^{- g V} \phi^I \right)
\, .
\end{equation}
The operator ${\cal K}$ (the kinetic term for matter) is usually called the
``Konishi operator''.

The conformal symmetry of ${\cal N} = 4$ SYM implies that two-point
functions of properly renormalised operators must take the form of
a power law ($N$ is the rank of the gauge group $SU(N)$)
\begin{equation}
\langle \bar O_r(x_2) \, O_r(x_1) \rangle \, = \,
\frac{c_r(g^2,N)}{\left(x_{12}^2 \right)^{\Delta_r(g^2,N)}} \, ,
\qquad x_{12} \, = x_1 - x_2 \, ; \qquad \langle \bar O_r(x_2) \,
O_s(x_1) \rangle \, = \, 0 \, : \quad \Delta_r \neq \Delta_s
\label{conf}
\end{equation}
where $r,s$ labels various operators. Quantum corrections can thus
affect the normalisation $c$ or lead to logarithms that sum into
the ``anomalous dimension'', i.e. the $g^2$ dependent part of
$\Delta$. Both equations only hold up to contact terms; in
dimensional regularisation or related schemes this means terms of
order $\epsilon$ before the regulator is sent to zero. Both ${\cal
O}$ and $O^{10}$ trivially satisfy the first equation in
(\ref{conf}) because quantum corrections are absent, in particular
$\Delta \, = \, 2, \, 3$, respectively. On the other hand, ${\cal
K}$ must be renormalised. By various arguments (first from the OPE \cite{kdim})
it is known that its dimension behaves like
\begin{equation}
\Delta_{\cal K} \, = \, 2 + 3 \, \alpha - 3 \, \alpha^2 + \ldots \, , \qquad \alpha \,
= \, \frac{ g^2 N}{4 \, \pi^2} \, .
\end{equation}

Curiously, $B$ as the ``descendant'' of ${\cal K}$ under the
classical equation of motion is not orthogonal to $O \, = \, F - 4
\, g \, B$. Moreover, from the explicit results below we would
calculate an anomalous dimension $\Delta_B \, = 3 + 3 \, \alpha -
9/2 \, \alpha^2 + \ldots$ and hence not the same as for ${\cal
K}$. This is in contradiction to supersymmetry: The ${\cal N} = 1$
supersymmetric version of the power law in (\ref{conf}) is
obtained by replacing $x_{12}$ by a supersymmetric line element.
Applying the differential operator from the equation of motion on
either end (thus $\bar D^2|_1 \, D^2|_2$) simply produces a box
operator (we illustrate this in Section 5 on the simpler case ${\cal O}$). The
dimension of the correct descendant is thus higher by one unit
whereas its anomalous part must agree.

According to \cite{konan,konishi} in the case of ${\cal K}$ the
equation of motion is modified in the quantum theory: The correct
descendant is
\begin{equation}
K \, := \, \frac{1}{12 \, g} \, \bar D_{\dot \alpha} \bar D^{\dot
\alpha} \, {\cal K} \, = \, B + \frac{g \, N}{32 \, \pi^2} \, F +
\ldots \, . \label{anoeq}
\end{equation}
This is an effect of the renormalisation of composite operators
and not a supersymmetry anomaly. When ${\cal K}$ is regularised by
point splitting it is in fact possible to derive the lowest order
$F$ admixture from the supersymmetry variation of a Wilson line
between the separated chiral fields in ${\cal K}$ \cite{konishi}.
The mixture $K$ is orthogonal to $O$ at order $g^1$, and we show
in the present article that it has the same anomalous dimension as
${\cal K}$ up to $O(g^4)$.

In theories without matter self-interaction the classical equation
of motion would send ${\cal K}$ to zero, which implies the
conservation of an axial current in the ${\cal N} = 1$ Grassmann
expansion of the operator. In the literature, the $F$ admixture
therefore has been coined the ``Konishi anomaly'' in analogy to
the standard axial anomaly. In several models without
superpotential the $F$ term does in fact not receive higher loop
corrections \cite{2345}. The more general case with matter self-interaction
is included in a similar statement in \cite{67}.

In \cite{EJSS} the question was taken up again in ${\cal N} = 4$
SYM by an analysis of the conformal properties of two-point
functions as stated in (\ref{conf}). Unfortunately, point
splitting is awkward to use at higher orders in perturbation
theory. Therefore the discussion was build around supersymmetric
dimensional reduction (SSDR) as a regulator \cite{siegel}, on the
expense of losing the direct derivation of the ``anomaly''. On
rather general grounds (the singularity structure and fractional
dimension of the correlators, supplemented by parametric
differentiation as in renormalisation group reasoning) it was
shown that the ``anomaly'' \emph{must} be affected by
renormalisation at least in the given scheme, in apparent
disagreement with \cite{67}.

To lend further support to the claim, we present a perturbative
treatment in SSDR at the second non-trivial
order.\footnote{Potential ambiguities in SSDR at very high orders
\cite{siegel2} do not play a role in this work.} As in \cite{EJSS}
we resort to the conformal properties (\ref{conf}) to resolve the
mixing. The programme is carried out from first principles: It is
shown that the two-point function of $O = F - 4 \, g \, B$ is
protected at $O(g^2)$ and $O(g^4)$. We then impose orthogonality
of
\begin{equation}
K \, = \, Z_B \, B + \frac{g \, N}{32 \pi^2} \, Z_F \, F
\end{equation}
to $O$ and the conformal form of the two-point function of $K$ at
the first two non-trivial orders in $g^2$. Of necessity, $Z_F$
contains singular higher loop corrections.

Upon rewriting $F = O + 4 \, g \, B$ and rescaling, the ``Konishi
descendant'' takes the form
\begin{equation}
\tilde K \, = \, \tilde Z_B \, B + \frac{g \, N}{32 \pi^2} \, \tilde Z_O \, O
\end{equation}
for which the Z-factors have been worked out in \cite{EJSS} in
closed form in terms of the anomalous part of the dimension:
\begin{eqnarray}
\tilde Z_B & = & \exp \left( \frac{1}{2 \, \epsilon} \int_0^\alpha
\frac{\gamma(\tau)}{\tau} \, d\tau \right) \, , \qquad
\gamma(\tau) \, = \, \gamma_1 \, \tau + \gamma_2 \, \tau^2 +
\ldots \label{closedZ} \\ \tilde Z_O & = & \frac{1}{\alpha} \left(
\omega(\alpha) - \tilde Z_B(\alpha, \epsilon) \int_0^\alpha
\omega(\tau) \,
\partial_\tau \left(\frac{1}{\tilde Z_B(\tau,\epsilon)} \right)
d\tau \right) \, , \qquad \omega(\tau) \, = \, \tau + \omega_1 \,
\tau + \omega_2 \, \tau^2 + \ldots \nonumber
\end{eqnarray}
Note that the two Z-factors are not proportional.
Apart from the anomalous dimension $\gamma(\alpha)$, graph
calculations can only determine the finite part $\omega(\alpha)$ of $\tilde
Z_O$. In terms of this data the poles in both Z-factors are given
by the formulae above. In particular, $\tilde Z_O$ is singular
even if $\omega_i \, = \, 0$.

The value $\omega_1 \, = \, - 3/4$ was given in \cite{EJSS}, below
we establish $\omega_2 \, = \, 7/8$. Moreover, up to $O(g^4)$ the
singular terms in the renormalisation factors obtained by direct
calculation coincide with the values expected from (\ref{closedZ}).

\newpage

\section{Loop corrections to the Konishi anomaly in SSDR}

In the next three sections of the article we work out the correlators
$\langle \bar B B \rangle$, $\langle \bar F F \rangle$ and $\langle \bar B F \rangle$
at the first three orders in perturbation theory. We use ${\cal N} = 1$ superfields
in supersymmetric Fermi-Feynman gauge. The regulator is supersymmetric dimensional
reduction. The hard part of the problem is $\langle \bar B F\rangle_{g^5}$ which contains
58 four-loop two-point superdiagrams. Breaking down the Grassmann algebra does not lead
to simple numerators. We have evaluated the integrals on the computer using the Laporta
algorithm, i.e. integration by parts. A sketch of the approach is given in Section 6.
The numerator reduction has been independently verified by A. Pak.

In this section we discuss the operator mixing problem between $B$ and $F$ drawing upon
the results, without touching upon the details of the graph calculation. The
renormalisation procedure laid out below is insensitive to division by regular functions
of $\epsilon$, the decrement of the space time dimension $D = 4 - 2 \epsilon$. In
equation (\ref{bits}) we have scaled down the correlators by
$\langle \bar B B \rangle_{g^0}$, so
\begin{equation}
BB_2 \, = \, \frac{\langle \bar B(x_2) B(x_1) \rangle_{g^2}}
{\langle \bar B(x_2) B(x_1) \rangle_{g^0}}
\end{equation}
etc. and we use the abbreviations (at the given orders there are no corrections
subleading in $N$)
\begin{equation}
X \, = \, x_{12}^2 \, \tilde \mu_x^2 \, , \qquad \alpha =
\frac{g^2 N}{4 \pi^2} \, .
\end{equation}
The vanishing of $\langle \bar B F \rangle_{g^1}$ is a trivial consequence of the
Feynman rules; there is no diagram.
\begin{eqnarray}
BB_0 \, = & 1 & \phantom{\frac{45}{4}} \label{bits} \\
BB_2 \, = & \frac{\alpha X^\epsilon}{\epsilon} & \left[ \, - 3 - 3
\, \epsilon - 9 \, \zeta(3)
\, \epsilon^2 \, + \ldots \, \right] \phantom{\frac{45}{4}} \nonumber \\
BB_4 \, = & \left(\frac{\alpha X^\epsilon}{\epsilon}\right)^2 &
\left[ \phm \frac{9}{2} + \frac{45}{4} \, \epsilon + \left(\frac{45}{4} + \frac{63}{2} \,
\zeta(3) \right) \, \epsilon^2 \, + \ldots \, \right] \nonumber \\
FF_0 \, = & 16 \, g^2 \, \frac{1}{\alpha X^\epsilon} & \left[ \, \phm 2 - 4
\, \epsilon + 2 \, \epsilon^2 \, + \ldots \, \right] \phantom{\frac{45}{4}} \nonumber \\
FF_2 \, = & 16 \, g^2 & \left[ - 1  - 12 \, \zeta(3)
\, \epsilon - \left( \frac{\pi^4}{5} - 54 \, \zeta(3) \right) \, \epsilon^2 \, +
\ldots \, \right] \nonumber \\
FF_4 \, = & 16 \, g^2 \, \frac{\alpha X^\epsilon}{\epsilon} & \left[ - \frac{3}{2} +
\frac{3}{2} \, \epsilon + \left(12 + 9 \, \zeta(3)  + \frac{75}{2} \, \zeta(5) \right)
\, \epsilon^2 \, + \ldots \, \right] \nonumber \\
BF_1 \, = & 0 & \phantom{\frac{45}{4}} \nonumber \\
BF_3 \, = & g \, \frac{\alpha X^\epsilon}{\epsilon} & \left[ \, - 9 - 3
\, \epsilon + 24
\, \epsilon^2 \, + \ldots \, \right] \phantom{\frac{45}{4}} \nonumber \\
BF_5 \, = & g \left(\frac{\alpha X^\epsilon}{\epsilon}\right)^2 &
\left[ \, \, \frac{33}{2} + 35 \, \epsilon + \left( - 3 + 90 \,
\zeta(3) \right) \, \epsilon^2 \, + \ldots \, \right] \nonumber
\end{eqnarray}
It follows immediately that the two-point function of $O \, = \, F- 4 \, g \, B$
is of order $\epsilon$ at $g^2$ and at $g^4$. In the limit $\epsilon \rightarrow 0$ these
are contact terms \cite{contact}. When $x_{12} \neq 0$ the one- and two-loop corrections
tend to zero --- in other words, this linear combination is protected. On the other hand,
protectedness at the first two orders strongly point towards all-loops protectedness.
Turning the argument around we might impose protectedness and view the absence of the
pole- and finite parts of
the two-point function of $F - 4 \, g \, B$ as a constraint relating the $\epsilon$
expansion of, say, $\langle \bar F F \rangle$ to that of the other correlators. In this
way the leading term in $FF_2$ and the two leading orders in
$FF_4$ must take the values given in the table. Moreover, $FF_2$ must not have a simple
pole and $FF_4$ must not have a double pole. A graph calculation merely yields
a consistency check.

The mixture $O$ has vanishing and hence well-defined anomalous dimension. Clearly
we can construct a second operator $K$ starting on $B$ that will have to be
orthogonal to $O$ if it has non-vanishing anomalous dimension. To lowest
order in $g$
\begin{equation}
K \, = \, B + \frac{g N}{32 \, \pi^2} F
\end{equation}
satisfies this constraint, but we also see that
\begin{equation}
\langle \bar K(2) \, O(1) \rangle_{g^1} \, = \, \langle \bar B(2) \, B(1)
\rangle_{g^0} \; 4 \, g \, \epsilon \log(x_{12}^2 \, \tilde \mu_X^2) \, + \ldots
\end{equation}
from which it is clear that we cannot expect the orthogonality constraint to hold beyond
$O(\epsilon^0)$. The bare two-point function of $K$ is divergent at $O(g^2)$ because
the pole in $BB_2$ cannot be compensated by the finite contribution
$FF_0$. We therefore renormalise as
\begin{eqnarray}
K & = & Z_B \, B \, + \,  \frac{g N}{32 \, \pi^2} \, Z_F \, F \, , \\
Z_B & = & 1 \, + \, \alpha \, \frac{b_{11}}{\epsilon}  \, + \, \alpha^2 \left(
\frac{b_{22}}{\epsilon^2} \, + \, \frac{b_{21}}{\epsilon} \right) \, + \ldots \, ,
\nonumber \\
Z_F & = & 1 \, + \, \alpha \, \left(\frac{f_{11}}{\epsilon}  \, + \, f_{10} \right) \, + \,
\alpha^2 \, \left(\frac{f_{22}}{\epsilon^2} \, + \, \frac{f_{21}}{\epsilon} \, +
f_{20} \right) \, + \ldots \nonumber \, .
\end{eqnarray}
Finite terms in $Z_B$ could be absorbed into an overall rescaling and a corresponding
change in $Z_F$; omitting these amounts to fixing the normalisation.
\begin{eqnarray}
\frac{\langle \bar K(2) \, K(1) \rangle_{g^2}}{\langle \bar B(2) \, B(1)
\rangle_{g^0}} & = & 2 \, \alpha \, \frac{b_{11}}{\epsilon} \, + \, BB_2 \, + \, \left(
\frac{g \, N}{32 \, \pi^2} \right)^2 \, FF_0 \\
& = & \alpha \left[ \left( 2 \, \frac{b_{11}}{\epsilon} \right) \, + \,
\left( - \frac{3}{\epsilon}
- 3 - 3 \log(X) \right) + \left( \frac{1}{2} \right) \, + O(\epsilon) \right] \nonumber
\end{eqnarray}
On the other hand, in the renormalised QFT conformal invariance implies that the
two-point function has the functional form
\begin{equation}
\frac{\langle \bar K(2) \, K(1) \rangle_{g^2}}{\langle \bar K(2) \, K(1)
\rangle_{g^0}} \, = \, \frac{1 + \alpha \, a_1 + \alpha^2 \, a_2 + \ldots}{X^{\alpha
\gamma_1 + \alpha^2 \, \gamma_2 + \ldots}} \, = \, 1 \, + \, \alpha \left(a_1 \, - \, \gamma_1
\, \log(X)\right) \, + \, O(\alpha^2) \, . \label{should2pt}
\end{equation}
In the first equation we can put $\epsilon$ to zero after adjusting $b_{11}$. It is then
possible to equate the last two lines. We learn
\begin{equation}
\gamma_1 \, = \, 2 \, b_{11} \, = \, 3 \, , \qquad a_1 \, = \, - \frac{5}{2} \, .
\end{equation}
Next, let us put
\begin{equation}
BF_3 \, = \, g \frac{\alpha X^\epsilon}{\epsilon} \, \left[ B_{31} - 3 \, \epsilon +
O(\epsilon^2) \, \right]
\end{equation}
for the moment and consider the orthogonality constraint at $O(g^3)$. Using $BB_0\, =
\, 1, \, BF_1 \, = \, 0$ we obtain
\begin{eqnarray}
\frac{\langle \bar K(2) \, O(1) \rangle_{g^3}}
{\langle \bar B(2) \, B(1) \rangle_{g^0}} & = & BF_3 - 4 \, g \, BB_2 -
4 \, g \, \alpha \, \frac{b_{11}}{\epsilon} + \frac{g \, N}{32 \, \pi^2} \, FF_2 +
\frac{g \, N}{32 \, \pi^2} \, \alpha \, \left(\frac{f_{11}}{\epsilon} + f_{10}\right)
\, FF_0 \\
& = & g \, \alpha \left[ \frac{1}{\epsilon} \left( B_{31} + 6 + 4 \,
f_{11} \right) + \log(X) \left( B_{31} + 12 - 4 \, f_{11} \right) +
\left( 7 - 8 \, f_{11} + 4 \, f_{10} \right) + O(\epsilon) \, \right] \, . \nonumber
\end{eqnarray}
This vanishes up to $O(\epsilon)$ if
\begin{equation}
B_{31} \, = \, - 9\, , \qquad f_{11} \, = \, \frac{3}{4} \, , \qquad f_{10} \, = \,
 - \frac{1}{4} \, .
\end{equation}
The singular term $f_{11}/\epsilon$ in $Z_F$ is
necessary, because we need to solve two different equations in order to eliminate the
pole and the logarithm. This fixes both the $f_{11}$ term and the pole term in $BF_3$.
According to (\ref{bits}) the graph calculation does indeed meet the requirement
$B_{31} \, = \, - 9$. The finite pieces in $BF_3, \, BB_2, \, FF_2$ and
$f_{11}$ times the subleading order in $FF_0$ determine the remaining coefficient
$f_{10} \, = \, - 1/4 \, \neq 0$. Hence in this scheme the Konishi anomaly is
affected by renormalisation.

Up to this point we reviewed the analysis of \cite{EJSS}. Next, we look at
\begin{eqnarray}
\frac{\langle \bar K(2) \, K(1) \rangle_{g^4}}
{\langle \bar B(2) \, B(1) \rangle_{g^0}} & = & BB_4 + \alpha \, \frac{2 \, b_{11}}{\epsilon} \,
BB_2 + \alpha^2 \, \left( \frac{b_{11}^2 + 2\, b_{22}}{\epsilon^2} + \frac{ 2 \,
b_{21}}{\epsilon} \right) + \\
&& + \, 2 \, \frac{g \, N}{32 \, \pi^2} \, BF_3 + \left( \frac{g \, N}{32 \, \pi^2} \right)^2 \, FF_2 + \, 2 \, \alpha \left(
\frac{f_{11}}{\epsilon} + f_{10} \right) \, FF_0 \nonumber
\end{eqnarray}
or by substituting the explicit formulae
\begin{eqnarray}
\frac{\langle \bar K(2) \, K(1) \rangle_{g^4}}
{\langle \bar B(2) \, B(1) \rangle_{g^0}} & = &
\alpha^2 \biggl[ \frac{1}{\epsilon^2} \left( B_{42} + 2 \, b_{22} - \frac{27}{4} \right) +
\frac{1}{\epsilon} \, \log(X) \left( 2 \, B_{42} - 9 \right) + (\log(X))^2 \left(
2 \, B_{42} - \frac{9}{2} \right) + \nonumber \\
&& \phantom{\alpha^2 \biggl[} + \frac{1}{\epsilon} \left( 2 \, b_{21} + \frac{3}{4}
\right) + \log(X) \, \frac{21}{2} + \left( \frac{17}{2} + \frac{9}{2} \, \zeta(3)
\right) + O(\epsilon) \, \biggr]
\end{eqnarray}
where we have put
\begin{equation}
BB_4 \, = \, \left(\frac{\alpha X^\epsilon}{\epsilon}\right)^2 \,
\left[ B_{42} + \frac{45}{4} \, \epsilon + \left(\frac{45}{4} + \frac{63}{2} \,
\zeta(3) \right) \, \epsilon^2 \, + \ldots \, \right]
\end{equation}
for now. The elimination of the singular terms leads to three conditions, one of which
we solve for $B_{42}$. Upon equating with (\ref{should2pt}) we find
\begin{equation}
B_{42} \, = \, \frac{9}{2} \, , \qquad b_{22} \, = \, \frac{9}{8} \, , \qquad
\gamma_2 \, = \, - 3 \, , \qquad b_{21} \, = \, - \frac{3}{8} \, , \qquad a_2 \, = \,
\frac{17}{2} + \frac{9}{2} \, \zeta(3) \, .
\end{equation}
Once again, the leading coefficient in $BB_4$ stated in (\ref{bits}) does in fact take
the right value.

It remains to analyse the orthogonality constraint at $O(g^5)$. The system is
overdetermined as before so that we start with
\begin{equation}
BF_5 \, = \, g \, \left(\frac{\alpha X^\epsilon}{\epsilon}\right)^2 \,
\left[ B_{52} + B_{51} \, \epsilon + \left(-3  + 90 \,
\zeta(3) \right) \, \epsilon^2 \, + \ldots \, \right] \, .
\end{equation}
(The pole part of $FF_4$ is already constrained by the protectedness of $O^{10}$).
We put in the other correlators from (\ref{bits}) and the coefficients of the Z-factors
already derived.
\begin{eqnarray}
\frac{\langle \bar K(2) \, O(1) \rangle_{g^5}}
{\langle \bar B(2) \, B(1) \rangle_{g^0}} & = &
g \, \alpha^2 \biggl[ \frac{1}{\epsilon^2} \left( B_{52} + 4 \, f_{22} - 18 \right) +
\frac{1}{\epsilon} \, \log(X) \left( 2 \, B_{52} - 4 \, f_{22} - \frac{63}{2} \right) +
\\ && \phantom{g \, \alpha^2 \biggl[} + (\log(X))^2 \left( 2 \, B_{52} + 2 \, f_{22} -
\frac{135}{4} \right) + \frac{1}{\epsilon} \left( B_{51} + 4 \, f_{21} - 8
\, f_{22} - 30 \right) + \nonumber \\ && \phantom{g \, \alpha^2 \biggl[}
+ \log(X) \, \left( 2 \, B_{51} - 4 \, f_{21} + 8 \, f_{22} - 75
\right) + \left(4 \, f_{20} - 8 \, f_{21} + 4 \, f_{22} - 7 \right) + O(\epsilon) \,
\biggr]
\nonumber
\end{eqnarray}
The correlator vanishes up to $O(\epsilon)$ if\footnote{We could have determined
$b_{22}$ from here, too.}
\begin{equation}
B_{52} \, = \, \frac{33}{2} \, , \qquad B_{51} \, = \, 35 \, , \qquad
f_{22} \, = \, \frac{3}{8} \, , \qquad f_{21} \, = \, - \frac{1}{2} \, , \qquad
f_{20} \, = \,  \frac{3}{8} \, .
\end{equation}
In conclusion, all three $O(\alpha^2)$ mixing coefficients in $Z_F$ are non-vanishing.
Interestingly, $\zeta(3)$ cancels from $f_{20}$.
The fact that the sum of the graphs in $\langle \bar B F \rangle_{g^5}$
reproduces $B_{52}, \, B_{51}$ is extremely non-trivial by looking at the orders
$1/\epsilon^3$ and $1/\epsilon^2$ in the explicit results in the
equations (\ref{usG0}) and (\ref{bigEq}). The constraints imposed by conformal
invariance after renormalisation give an excellent test of the graph
calculation.

To make touch with the discussion in \cite{EJSS}, we write
$F \, = \, O + 4 \, g \, B$ and divide $K$ by $1 +
\alpha/2 - \alpha^2/8$ in order to eliminate finite contributions from the
shifted renormalisation factor $Z_B + \alpha Z_F/2$ which now multiplies $B$.
The renormalised operator mixture fits the general form
\begin{eqnarray}
\tilde K & = & \tilde Z_B \, B + \frac{g \, N}{32 \, \pi^2} \, \tilde Z_O \, O \\
\tilde Z_B & = & 1 + \alpha \frac{\gamma_1}{2 \, \epsilon} + \alpha^2 \left(
\frac{\gamma_1^2}{8 \, \epsilon^2} + \frac{\gamma_2}{4 \, \epsilon} \right) + \ldots
  \nonumber \\
\tilde Z_O & = & 1 + \alpha \left( \frac{\gamma_1}{4 \, \epsilon} + \omega_1 \right)
+ \alpha^2 \left(\frac{\gamma_1^2}{24 \, \epsilon^2} + \frac{\gamma_2 +
\omega_1 \gamma_1}{6 \, \epsilon} + \omega_2 \right) + \ldots \nonumber
\end{eqnarray}
predicted by (\ref{closedZ}), with
\begin{equation}
\gamma_1 \, = \, 3 \, , \qquad \gamma_2 \, = \, - 3 \, , \qquad
\omega_1 \, = \, - \frac{3}{4} \, , \qquad \omega_2 \, = \, \frac{7}{8} \, .
\end{equation}
Our effort shows that the renormalisation scheme developed in \cite{EJSS} is operational
also at the second order in $\alpha$, and it yields the previously unknown number
$\omega_2$ which could be called the two-loop Konishi anomaly.

\section{$\langle\bar B F\rangle_{g^5}$}
The complete set of superdiagrams is displayed in Figure 1 and Figure 6.
The diagrams in Figure 1 are special in that they have two cubic non-abelian
vertices or one non-abelian four-vertex. They all turn out to be derived
topologies\footnote{topologies that arise by cancelling a line} of the
first diagram $G_0$ so that they can conveniently be summed into one
effective numerator $\tilde G_0$.

\vskip 0.2 cm
\hskip -0.5 cm
\begin{minipage}{\textwidth}
$G_0$ \\
\includegraphics[width = 0.796 \textwidth]{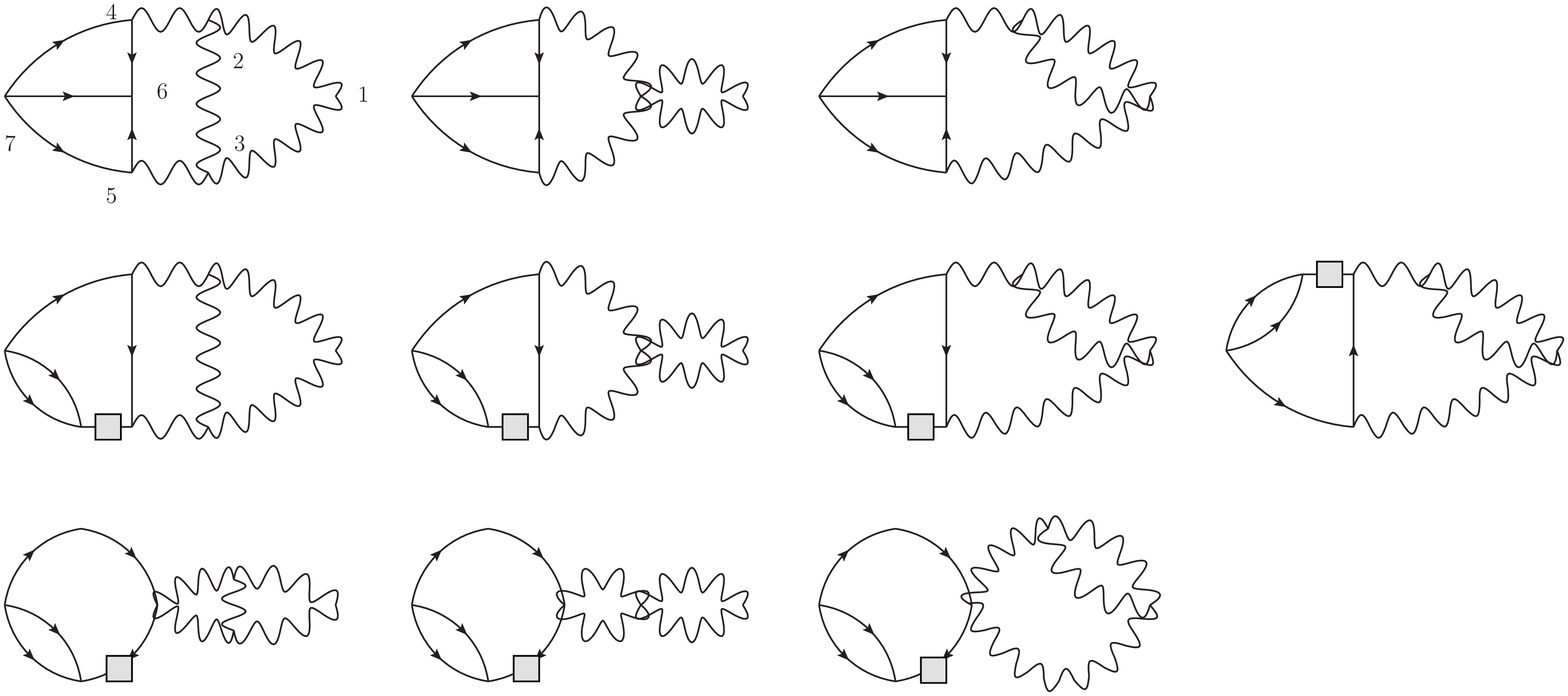}
\end{minipage}
\phantom{W} \hskip 5 cm \textbf{Figure 1}
\vskip 0.2 cm

Suppose that we start by Wick contracting only the matter part of the diagrams,
i.e. the operator $\bar B(x_7)$ on the cubic chiral vertex at point $6$ and
the respective matter/YM vertices at point 4 and 5, where present. There are
three ``matter parts'':

\vskip 0.3 cm
\hskip - 0.5 cm
\begin{minipage}{\textwidth}
$M_1$ \hskip 2.74 cm $M_2$ \hskip 2.74 cm $M_3$ \\
\includegraphics[width = 0.591 \textwidth]{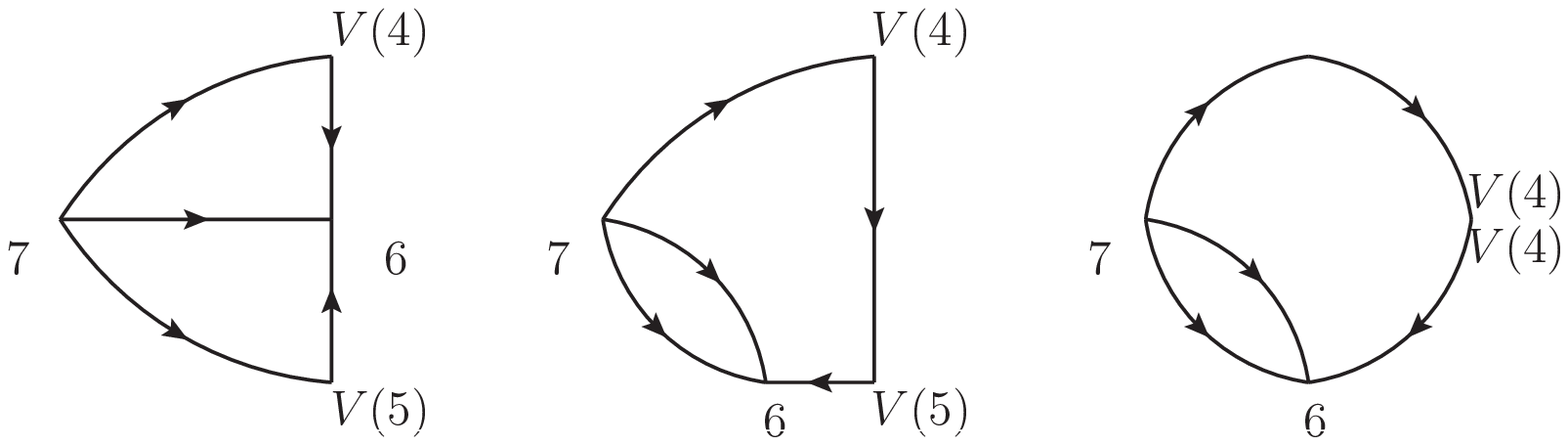}
\end{minipage}
\vskip - 0.25 cm
\phantom{W} \hskip 5 cm \textbf{Figure 2}
\vskip 0.2 cm
The matter line has an arrow pointing from the antichiral to the
chiral end. One way of writing this propagator is
\begin{equation}
\langle \bar \phi(i) \, \phi(j) \rangle \, = \, D . D \, \bar D . \bar D |_i \,
\Pi_{ij} \, , \qquad D . D \, = \, - \frac{1}{4} D^\alpha D_\alpha \, ,
\qquad \bar D . \bar D \, = \, - \frac{1}{4} \bar D_{\dot \alpha}
\bar D^{\dot \alpha} \, . \label{matprop1}
\end{equation}
Here $\phi(i) = \phi(x_i,\theta_i,\bar \theta_i)$, and
\begin{equation}
\Pi_{ij} = \frac{\delta(\theta_{ij}) \, \delta(\bar \theta_{ij})}{- 4 \pi^2
x_{ij}^2} \, \qquad \delta(\theta_{ij}) \, = \, \theta_{ij}^\alpha
\theta_{ij \alpha} \, , \qquad \delta(\bar \theta_{ij}) \, = \,
\bar \theta_{ij \dot \alpha} \bar \theta_{ij}^{\dot \alpha} \label{piprop}
\end{equation}
where $x_{ij}, \theta_{ij} \bar \theta_{ij}$ denote coordinate differences. The Yang-Mills
propagator is $\langle V(i) \, V(j) \rangle \, = \, - \Pi_{ij}$.

For this class of graphs we will use ``$D$-algebra'' to derive the numerators.
The tools are partial integration, the commutation relation
$\{D_\alpha, \bar D_{\dot \alpha}\} \, = \, - 2 i \partial_{\alpha \dot\alpha}$
and the shift rule
\begin{equation}
D_i \, \Pi_{ij} \, = \, - D_j \Pi_{ij} \, .
\end{equation}
The idea is to free all but one line of each loop from spinor derivatives.
Due to the shift identity and the possibility of reordering via the commutation
relations, we may reduce to spacetime derivatives and maximally two chiral and
two antichiral spinor derivatives on each $\Pi_{ij}$. On the last line in
each loop we need to keep only terms with all four derivatives, since
\begin{equation}
\delta(\theta_{ij}) \, \delta(\bar \theta_{ij}) \, \left( D.D \bar D. \bar D \,
\delta(\theta_{ij}) \, \delta(\bar \theta_{ij}) \right) \, = \,
\delta(\theta_{ij}) \, \delta(\bar \theta_{ij}) \label{DbetweenDeltas}
\end{equation}
while less than four derivatives between two $\delta$-functions vanish.

Let us evaluate matter part $M_1$ with this technique: We absorb the
$\bar D . \bar D$ of the 7-6 line into the measure at the chiral vertex 6
and partially integrate the $D . D$ away from it onto the 4-6 and the 5-6
lines. We then shift all spinor derivatives to points 4 and 5 respectively.
Last, we partially integrate the spinor derivatives away from the 4-6 line,
onto V(4) and the 7-4 line, likewise at point 5. The result is
\begin{eqnarray}
\frac{M_1}{- 6 N^2 g^3} & = & \mathrm{Tr}\left[ (D.D \, V(4)) \,
V(5) \right] \, \square_{46} \, \square_{57} \, + \,
\mathrm{Tr}\left[ (D.D \, V(4)) \, (\bar D_{\dot \alpha} D_\alpha
\, V(5)) \right] \, \square_{46} \, \frac{i}{2} \,
\partial_{57}^{\dot \alpha \alpha} \, - \nonumber \\
&& \mathrm{Tr}\left[ (D.D \, V(4)) \, (\bar D . \bar D \, D . D \,
V(5)) \right] \, \square_{46} \, + \nonumber \\ &&
\mathrm{Tr}\left[ V(4) \, (D.D \, V(5)) \right] \, \square_{56} \,
\square_{47} \, + \, \mathrm{Tr}\left[ (\bar D_{\dot \alpha}
D_\alpha \, V(4)) (D.D \, V(5)) \right] \, \square_{56} \,
\frac{i}{2} \,
\partial_{47}^{\dot \alpha \alpha} \, - \nonumber
\\ && \mathrm{Tr}\left[ (\bar D . \bar D \, D . D \, V(4)) (D.D \,
V(5)) \right] \, \square_{56} \, - \nonumber \\ &&
\mathrm{Tr}\left[ (D^\alpha \, V(4)) \, (D_\alpha V(5)) \right] \,
\frac{1}{2} \, \square_{46} \, \square_{56} \, -  \nonumber \\ &&
\mathrm{Tr}\left[ (D_\alpha \, V(4)) \, (D.D \, \bar D_{\dot
\alpha} \, V(5)) \right] \,  \square_{46} \, \frac{i}{2} \,
\partial_{56}^{\dot \alpha \alpha} \, + \nonumber \\
&& \mathrm{Tr}\left[ (D.D \, \bar D_{\dot \alpha} \, V(4)) \,
(D_\alpha \, V(5)) \right] \,  \square_{56} \, \frac{i}{2} \,
\partial_{46}^{\dot \alpha \alpha} \, - \nonumber \\ && \mathrm{Tr}\left[
(D.D \, \bar D_{\dot \alpha} \, V(4)) \, (D.D \, \bar D^{\dot
\beta} \, V(5)) \right] \,  \frac{1}{2} \, \partial_{46}^{\dot
\alpha \alpha} \, \partial_{56 \alpha \dot \beta}
\end{eqnarray}
In the last formula $\partial_{ij}$ only acts on the $i-j$ line, at point $i$.
For notational convenience we have omitted $\Pi_{46} \Pi_{47} \Pi_{56} \Pi_{57}
\Pi_{67}$ and the integrations. Similarly,
\begin{eqnarray}
\frac{M_1}{6 N^2 g^3} & = & \mathrm{Tr}\left[ V(4) \, (D.D \, V(5)) \right]
\, \square_{56} \, \square_{47} \, + \, \mathrm{Tr}\left[ (\bar D_{\dot \alpha}
D_\alpha \, V(4)) (D.D \, V(5)) \right] \, \square_{56} \,
\frac{i}{2} \, \partial_{47}^{\dot \alpha \alpha} \, - \nonumber
\\ && \mathrm{Tr}\left[ (\bar D . \bar D \, D . D \, V(4)) (D.D \,
V(5)) \right] \, \square_{56} \, + \nonumber \\ &&
\mathrm{Tr}\left[ (D.D \, V(4)) \, V(5) \right]
\, \square_{56} \, \square_{45} \, - \,
\mathrm{Tr}\left[ (D^\alpha \, V(4)) \, (D_\alpha \, V(5)) \right]
\, \frac{1}{2} \, \square_{56} \, \square_{45} \, + \nonumber \\
&& \mathrm{Tr}\left[ (D.D \, \bar D_{\dot \alpha} \, V(4)) \,
(D_\alpha \, V(5)) \right] \,  \square_{56} \, \frac{i}{2} \,
\partial_{45}^{\dot \alpha \alpha}
\end{eqnarray}
Next, $\square_{56}$ leads to the same derived topology in $M_1$ and $M_2$
(by shrinking the 5-6 line to a point) so that the
``numerators'' can simply be added:
\begin{eqnarray}
\frac{M_1 \, + \, M_2 \, + (M_2 : (4 \leftrightarrow 5))}{6 N^2 g^3} & = &
\mathrm{Tr} \left[ \bar \lambda_{\mathrm{lin} \, \dot \alpha}(4) \,
\bar \lambda_{\mathrm{lin}}^{\dot \beta}(5) \right] \, \frac{1}{2} \,
\partial_{46}^{\dot \alpha \alpha} \, \partial_{56 \alpha \dot \beta} \, + \,
\bigl(\mathrm{Tr} \left[ (D . D \, V(4)) \, V(5) \right] \, -
\label{M12}\\
&& \frac{1}{2} \mathrm{Tr} \left[ (D^\alpha \, V(4)) \,
(D_\alpha \, V(5)) \right] + \mathrm{Tr} \left[ V(4) \, (D . D \, V(5)) \right]
\bigr) \, \square_{46} \, \square_{56} \nonumber
\end{eqnarray}
with the linear part of the antichiral physical fermion in the gauge multiplet
$\bar \lambda_{lin}^{\dot \alpha} \, = \, D . D \, \bar D^{\dot \alpha} \, V$.
On the other hand:
\begin{equation}
\frac{M_3}{6 N^2 g^3} \, = \, - D . D \, \mathrm{Tr} \left[ V(4) V(4) \right] \,
\square_{46}
\end{equation}
Due to Greens' function equation $\square_{56} \, \Pi_{56} \, = \, +
\delta^8(z_{56})$ this cancels the $\square_{46} \square_{56}$ terms in
(\ref{M12}). In Figure 1 we can therefore work with one effective matter part
\begin{equation}
\tilde M_1 \, = \, 3 N^2 g^3 \, \mathrm{Tr} \left[ \bar \lambda_{\mathrm{lin} \,
\dot \alpha}(4) \, \bar \lambda_{\mathrm{lin}}^{\dot \beta}(5) \right] \,
\partial_{46}^{\dot \alpha \alpha} \, \partial_{56 \alpha \dot \beta} \, .
\label{effectiveMat}
\end{equation}
Note that putting $\theta_7 \, = \bar \theta_7 = 0$ also selects the
$\theta = \bar \theta = 0$ component of $\bar \lambda_{\mathrm{lin}}$ at both
open ends.

To show consistency with \cite{EJSS} we recompute the leading
$O(g^3)$ contribution to $\langle \bar B F\rangle$. With
\begin{equation}
F \, = \, \mathrm{Tr} \left[ \lambda^\alpha \, \lambda_{\alpha} \right] \,
\qquad \lambda^\alpha \, = \, - \frac{1}{g} \, \bar D . \bar D \, e^{ g V}
\, D^\alpha \, e^{-g V} \nonumber
\end{equation}
we find
\begin{equation}
\langle \bar B(7) \, F(1) \rangle_{g^3} \, = \, - 24 \, g^3 N^2 (N^2-1) \,
\partial_{46}^{\dot \alpha \alpha} \, \partial_{56 \, \alpha \dot \beta} \,
\partial_{15}^{\dot \beta \beta} \, \partial_{14 \, \beta \dot \alpha} \, .
\end{equation}
Once again, for notational convenience we only gave the numerator and
omitted the integral itself.

Throughout the paper we employ supersymmetric Fermi-Feynman gauge.
The last result would in fact have been immediate in Wess-Zumino (WZ) Fermi-Feynman
gauge. However, the D-algebra can be done with the aid of a
computer programme so that the non-supersymmetric gauge does not offer any
advantage for the $g^5$ part of the correlator.

\hskip - 0.5 cm
\begin{minipage}{\textwidth}
\includegraphics[width = 0.54 \textwidth]{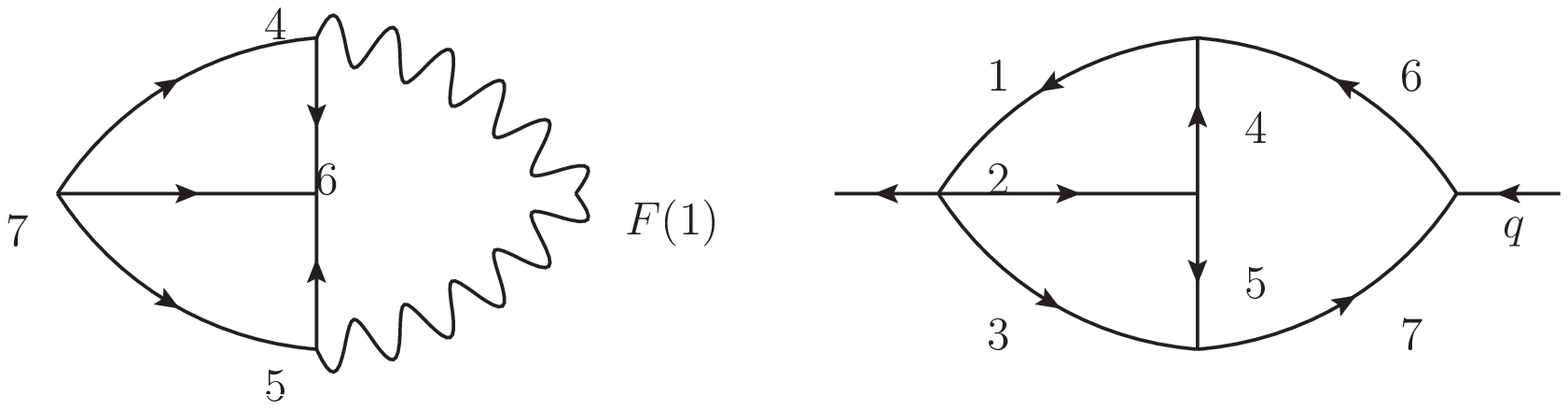}
\end{minipage}
\phantom{W} \hskip 5 cm \textbf{Figure 3}
\vskip 0.2 cm

The first picture in Figure 3 displays the
effective matter part $\tilde M_1$ contracted onto $F(1)$, or equivalently
the one graph in $\langle \bar B \, F \rangle_{g^3}$ in WZ gauge; here the
numbers label the spacetime points. The second picture reproduces the momenta
assignment from the discussion
of three-loop p-integrals in \cite{Mincer}, where this topology is called
\textbf{BU}. Fourier transform leads to the identification
\begin{equation}
\partial_{14} \rightarrow i \, p_6 \, , \qquad
\partial_{15} \rightarrow - i \, p_7 \, , \qquad
\partial_{46} \rightarrow - i \, p_4 \, , \qquad
\partial_{56} \rightarrow - i \, p_5 \, . \qquad
\end{equation}
The negative sign of the five matter propagators accounts for an extra minus,
so that in momentum space
\begin{eqnarray}
\langle \bar B \, F \rangle_{g^3} & = & - 48 \, g^3 N^2 (N^2-1) \left[
(p_4.p_5) (p_6.p_7) - (p_4.p_7) (p_5.p_6) + (p_4.p_6) (p_5.p_7) \right] \\
& = & - \frac{12 \, g^3 N^2 (N^2-1)}{(4 \pi)^6}
\left[ \frac{1}{\epsilon^2} \, + \, \frac{28}{3} \frac{1}{\epsilon} \, + \,
\frac{166}{3} \, + \, O(\epsilon^1) \right] \, q^2 \, (\frac{q^2}{\tilde \mu^2})^
{- 3 \epsilon}
\, . \nonumber
\end{eqnarray}
The second line shows the result of \emph{Mincer} in its $\overline{MS}$
mode.\footnote{The definition of the mass-scale $\tilde \mu$ is given before
equation (\ref{usG0}).}

After this digression let us return to the diagrams of Figure 1. Since
combinatorics is linear, the collection of ten supergraphs can be viewed as
the effective matter part contracted onto the pure Yang-Mills correlator
\begin{equation}
\tilde Y_1 \, = \, \langle \bar \lambda_{\dot \alpha \, \mathrm{lin} }^a(4) \, \bar
\lambda_{\mathrm{lin}}^{\dot \beta b }(5) \, F(1) \rangle_{g^2}
\end{equation}
where $a,b$ are colour indices. The latter contains the following graphs:
\vskip 0.3 cm
\hskip - 0.5 cm
\begin{minipage}{\textwidth}
$Y_1$ \hskip 3.65 cm $Y_2$ \hskip 3.65 cm $Y_3$ \hskip 3.65 cm $Y_4$ \\
\includegraphics[width = \textwidth]{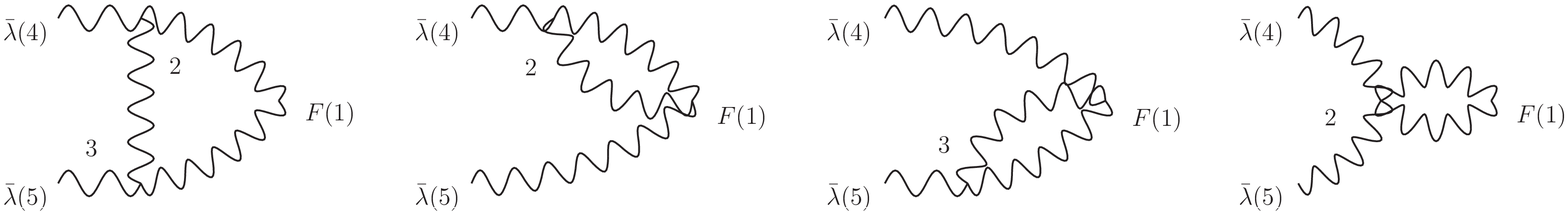}
\end{minipage}
\phantom{W} \hskip 5 cm \textbf{Figure 4}
\vskip 0.2 cm

The D-algebra for this sector is cumbersome; as an example we sketch
a strategy for the first diagram $Y_1$. There are three derivatives on each
outer leg and four at each vertex. We resorted to a \emph{Mathematica}
script for Wick contraction and superspace algebra. One may choose to shift
the derivatives to point 2 and 3, respectively, and then clear the 1-2
and 1-3 lines by partial integration. The derivatives on the 2-3 line are
subsequently shifted to one end and reordered. Due to formula
(\ref{DbetweenDeltas}) only terms with $D.D \, \bar D . \bar D$ need to be
kept. Next, we regroup the derivatives on the 2-4 and the 3-5 legs.
Upon putting the outer theta variables to zero
only terms with four spinor derivatives on both the 2-4 and the
3-5 legs survive.

In a similar way,
\begin{eqnarray}
\frac{Y_2}{2 N g^2 \, \delta^{ab}} & = &  \frac{7}{2} \left( \delta_{\dot \alpha}^{\dot
\beta} \, (\square_1 - \square_{15} - \square_{24}) + (\partial_{15}^{\dot \beta \beta} \,
\partial_{24 \beta \dot \alpha} - \partial_{24}^{\dot \beta \beta} \,
\partial_{15 \beta \dot \alpha}) \right) - 3 \, \delta_{\dot \alpha}^{\dot \beta} \,
\square_{24} \, , \\
\frac{Y_4}{2 N g^2 \, \delta^{ab}} & = &  \delta_{\dot \alpha}^{\dot
\beta} \, (\square_1 +\frac{2}{3} \, \square_{15} + \frac{2}{3} \, \square_{24}) +
(\partial_{24}^{\dot \beta \beta} \,
\partial_{25 \beta \dot \alpha} - \partial_{25}^{\dot \beta \beta} \,
\partial_{24 \beta \dot \alpha}) \nonumber
\end{eqnarray}
where $\partial_1$ is a total derivative w.r.t. $x_1$.
Clearly $Y_2,\, Y_3, \, Y_4$ arise from $Y_1$ by $\square_{13}, \, \square_{12}, \,
\square_{23}$, respectively. Here the Green's function equation brings in an additional
minus sign because $\langle V(i) V(j) \rangle = - \Pi_{ij}$. Upon adding these
contributions into $Y_1$:
\begin{eqnarray}
\frac{\tilde Y_1}{2 N g^2 \delta^{ab}} & = & 4 \, \square_1
(\partial_{24}^{\dot \beta \beta}
\, \partial_{35 \beta \dot \alpha} - \partial_{35}^{\dot \beta \beta} \,
\partial_{24 \beta \dot \alpha}) - (\square_{24} + \square_{35})
(\partial_{1}^{\dot \beta \beta} \, \partial_{23 \beta \dot \alpha} -
\partial_{23}^{\dot \beta \beta} \, \partial_{1 \beta \dot \alpha})
\label{effectiveYM} \\
&& - 4 \, \square_1 (\square_1 - \square_{24} - \square_{35}) -
(\square_{24} - \square_{35})^2 + (\square_{12} \square_{35} + \square_{13} \square_{24})
\nonumber \\
&& + (\square_{12} \square_{24} + \square_{13} \square_{35}) - \frac{2}{3} (\square_{24} +
\square_{35}) \square_{23} \nonumber
\end{eqnarray}
As another cross-check we derive $\langle \bar F(4) \, F(1)
\rangle_{g^2}$ from these formulae:
\begin{equation}
\delta_{\dot \beta}^{\dot \alpha} \, \delta^{ab} \, \tilde Y_1(5 \rightarrow 4) \, = \,
2 g^2 N (N^2-1) \left[
- 8 \, \square_1^2 + 7 \, \square_1 (\square_{24} + \square_{34}) + 3 \,
(\square_{12} \square_{34} + \square_{13} \square_{24}) \right]
\end{equation}
When identifying the points tadpole terms were put to zero. Finally,
the mirror image of $Y_2, \, Y_3$ arising from the cubic part of $\bar F(4)|_g$ and the
tree-like contribution from  $\langle \bar F(4)|_g \, F(1)|_g
\rangle_{g^2}$ have to be added. The result is
\begin{equation}
\langle \bar F(4) \, F(1) \rangle_{g^2} \, = \, 32 (N^2-1) \, g^2 N \,
(\square_{12} \square_{34} - \frac{1}{2} \, \square_1^2) \label{ffg2}
\end{equation}
consistent with the results in \cite{EJSS} for the $SU(4)$ component $\check F \, = \,
\mathrm{Tr} ((\nabla^\alpha \Phi^1) (\nabla_\alpha \Phi^1))$ of $F^{10}$.
According to (\ref{h1h2}) in the next section, the second term in the last
formula is a contact contribution proportional to $\zeta(3)$. The two box operators
in the first term break the integrations and we find
$- 32 \, g^2 N (N^2-1) \, \langle \bar \phi(4) \phi(1) \rangle^3$ due to the opposite sign
of the matter
propagator. In the combination $F - 4 \, g \, B$ this is cancelled
by $16 \, g^2 \langle \bar B(4) \, B(1) \rangle_{g^0}$ as required by protectedness.

By putting formulae (\ref{effectiveMat}) and (\ref{effectiveYM}) together we obtain the
desired effective numerator $\tilde G_0$ for the first topology in Figure 1. Graph $G_0$
is of topology \textbf{C} in the list in Figure 5 (the numbers in the figure label the
momenta $p_1 \ldots p_{10}$).

\vskip 0.3 cm
\hskip - 0.5 cm
\begin{minipage}{\textwidth}
$\mathbf{A}$ \hskip 3.65 cm $\mathbf{B}$ \hskip 3.65 cm
$\mathbf{C}$ \hskip 3.65 cm $\mathbf{D}$ \\
\includegraphics[width = \textwidth]{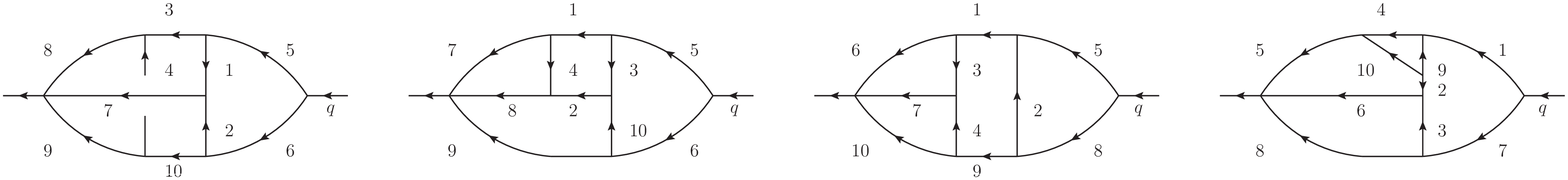}
\end{minipage}
\phantom{W} \hskip 5 cm \textbf{Figure 5}
\vskip 0.2 cm

In Section 6 we comment on the realisation of the Laporta algorithm
\cite{Laporta} with which we evaluated such four-loop diagrams with rather
general six-derivative numerators. In topology \textbf{C} we have chosen the
momenta $q,p_1,p_2,p_3,p_4$ as independent. To calculate $\tilde G_0$ we
identify
\begin{equation}
\partial_1 \, \rightarrow \, i \, q \, , \qquad
\partial_{24} \, \rightarrow \, i \, p_1 \, , \qquad
\partial_{23} \, \rightarrow \, - i \, p_2 \, , \qquad
\partial_{46} \, \rightarrow \, i \, p_3 \, , \qquad
\partial_{56} \, \rightarrow \, i \, p_4 \, .
\end{equation}
Between these one may form ten dot products. In total there are ten interior
lines, so that six of the dot products may be eliminated in favour of the four others
and the squares of the dependent momenta. In terms of $p_2.p_3, \, p_2.p_4, \, p_3.q,
\, p_4.q$ and $p_1^2 \ldots p_{10}^2, \, q^2$
the effective numerator has 47 terms. Flip symmetries of the original topology
and derived diagrams could be used to simplify.
\begin{equation}
\tilde G_0 \, = \, 12 \, g^5 N^3 (N^2-1) \left[ 8 \, q^2 \, p_3.q \, p_4.q \,
- 2 \, (p_2.p_4 \, p_3.q - p_2.p_3 \, p_4.q)
(p_1^2 \, + \, p_9^2) + (42 \; \mathrm{terms}) \, \right] \label{cnum}
\end{equation}
(The overall negative sign from the odd number of matter propagators has been taken
into account.) The omitted terms all have at least two distinct complete squares. Each
$p_i^2, \, i = 1 \ldots 10$ cancels a line in the denominator and thus leads to a
simplification within the Laporta approach (although $p_i^4$ is generally not better than $p_i^2$).
The first term in the square brackets is thus the hardest to analyse; an algorithm
that solves this problem can also calculate the other pieces.

In the $\overline{MS}$ scheme fractional powers of 4 and $\pi$ and the Euler constant
are absorbed into the mass scale by putting $\bar \mu^2 \, = \, \mu^2 \, 4 \pi e^{-\gamma_E}$.
One may further scale away $\zeta(2)$ by the redefinition $\tilde \mu^{
2 \epsilon} \, = \, \bar \mu^{2 \epsilon}
(1 - \epsilon^2 \pi^2/12 + O(\epsilon^4))$. In this convention our result for
the sum of the diagrams in Figure 1 takes the form
\begin{equation}
\tilde G_0 \, = \, - \frac{12 \, g^5 N^3 (N^2-1)}{(4 \pi)^8} \left[\frac{19}{72}
\frac{1}{\epsilon^3}+\frac{53}{16} \frac{1}{\epsilon^2}+
\left(\frac{7957}{288} - \frac{53}{4} \zeta(3) \right)
\frac{1}{\epsilon} + O(\epsilon^0) \right] q^2 \, \left(\frac{q^2}{\tilde \mu^2}\right)^{-4
\epsilon} \, . \label{usG0}
\end{equation}
At the first glimpse not all diagrams in Figure 1 can in fact be drawn into topology \textbf{C}.
We have shown above that the numerators of the matter parts $M_2, \, M_3$ have d'Alembert
operators acting onto the 5-6 and 4-6 lines, respectively, common to all terms. These are marked
by grey boxes in the graphs of Figure 1. Contracting the corresponding
lines yields topologies that can also be derived from \textbf{C}.

We have indicated such ``topology changing'' box operators on the diagrams of Figure 6, too.
For instance, graphs $G_{31}$ and $G_{32}$ (the one-loop correction to the chiral vertex
in matter part $M_1$) have originally neither of the four topologies
\textbf{A}, \textbf{B}, \textbf{C}, \textbf{D}.
The grey box operators are positioned such that $G_{31}$ can be obtained from \textbf{B} by
a factor $p_1^2$, and $G_{32}$ from \textbf{A} by $p_7^2$ or equivalently from \textbf{D} by
$p_5^2$.

Figure 6 contains several classes of diagrams that could be summed prior to
evaluation like the graphs of Figure 1. The graphs $G_{16}
\ldots G_{20}$, $G_{21} \ldots G_{25}$, $G_{26} \ldots G_{30}$ and $G_{31},G_{32}$; $G_{33}, \, G_{34}$
and finally $G_{40},\, G_{41}$ are all vertex corrections relating to the matter parts $M_1,M_2,M_3$
contracted onto $F(1)$. Even in the most complicated graphs $G_{16}, \, G_{19}, \, G_{20}$ (type
\textbf{D}) there is at least one square of an interior momentum in each numerator term, hence
the complexity is lower than for $\tilde G_0$. There is no calculational disadvantage in individually
treating the diagrams, so that we separately list the contribution of every supergraph in order to
make the whole calculation more verifiable.

The longest numerators for the \textbf{A}, \textbf{B} topologies arise from the graphs $G_3$ and
$G_{10}$ in Figure 6 with 134 and 183 numerator terms in our momenta assignments, respectively.
These as well as the numerators of $G_1, \, G_6, \, G_7$ contain terms with no
squared momentum. Yet, in all such
terms at least two of the three non-trivial dot products involve the outer momentum $q$. The
numerator reduction can then still be accomplished with a slightly restricted matrix; terms with
one or no $q$ are not needed in the ansatz since $q$ cannot be made to disappear by differentiation
w.r.t. loop momenta.

In our way of organising the Laporta algorithm by stepwise elimination of lines (see Section
6), topology \textbf{A} is the hardest case because it contains only the triangle between
$p_4,p_8,p_9$. Using the triangle rule (\ref{triagRule}) we immediately fall upon two derived topologies
that must be further reduced by the Laporta ansatz. In topology \textbf{B} we use
(\ref{triagRule}) on the triangle consisting of $p_4,p_7,p_8$,
which produces a new triangle in every term. The step to the eight-propagator level is therefore
more direct. Topologies \textbf{C} and \textbf{D} both contain two disjoint triangles.

Graph $G_4$ is in the same way a subgraph of $G_3$ as the matter part $M_2$ belongs to $M_1$;
likewise, $G_{19}$ and $G_{29}$ go together, and $G_{10}$ and $G_{24}$. On the other hand, $G_{24}$
naturally belongs to a class of vertex corrections, while $G_{10}$ does not. It is thus not
clear how to assemble all the superdiagrams in Figure 6 into well-defined classes; likewise
$G_{11}$ and its subcase $G_{12}$ could be attributed to either of the three pairs with one
non-abelian vertex.

The ``D-algebra'' technique is not convenient in diagrams with too many matter lines, because
according to (\ref{matprop1}) every such propagator brings in four spinor derivatives.
In the diagrams of Figure 6 we have rather
evaluated the Yang-Mills part (in $G_1$ and similar diagrams the two lines attached to $F(1)$,
in $G_3$ and $G_{11}$ etc. the complete gluon part) by D-algebra and then taken
the remaining spinor derivatives to obtain the Grassmann expansion at $\theta_1 = \bar \theta_1 = 0$.
The alternative form of the matter propagator
\begin{equation}
\langle \bar \phi(i) \, \phi(j) \rangle \, = \, e^{- i ((\theta_i \partial_{x_i} \bar \theta_i) +
(\theta_j \partial_{x_i} \bar \theta_j) - 2 (\theta_j \partial_{x_i} \bar \theta_i))} \,
\frac{1}{- 4 \pi^2 x_{ij}^2}
\end{equation}
easily lends itself to Grassmann expansion. The complete numerators are found by taking the product
with the expansion of the Yang-Mills parts and integrating out the spinor variables.\footnote{
We have computed $\tilde G_0$ also by diagram-wise evaluation using this technique.}

Finally, from the sum of $\tilde G_0$ and $G_1 \ldots G_{48}$ of Figure 6 / equation (\ref{bigEq})
we find
\begin{equation}
\langle\bar B(7) \, F(1) \rangle_{g^5} \, = \,  \frac{12 \, g^5 N^3 (N^2-1)}{(4 \pi)^8} \left[
\frac{11}{2}
\frac{1}{\epsilon^3}+\frac{899}{12} \frac{1}{\epsilon^2}+
\left(\frac{15043}{24} + 30 \, \zeta(3) \right)
\frac{1}{\epsilon} + O(\epsilon^0)
\right] q^2 \, \left(\frac{q^2}{\tilde \mu^2}\right)^{-4
\epsilon}
\end{equation}
in momentum space. The backward Fourier transform\footnote{The forward transform lacks the
factor $1/(2 \pi)^D$ and has the roles of $q,x$ and the sign in the exponent exchanged.}
\begin{equation}
\frac{1}{(2 \pi)^D} \int d^Dx \, e^{- i \, q . x} \, \frac{1}{(q^2)^\alpha} \, = \,
\frac{1}{4^\alpha \pi^{\frac{D}{2}}} \,
\frac{\Gamma(\frac{D}{2}-\alpha)}{\Gamma(\alpha)} \, \frac{1}{(x^2)^{\frac{D}{2} - \alpha}}
\end{equation}
yields ($\tilde \mu_x^{2 \epsilon} \, = \, \left(\mu^2 \pi e^{\gamma_E}\right)^\epsilon
\left(1 + \epsilon^2 \pi^2 / 12 + O(\epsilon^4)\right)$ in configuration space)
\begin{equation}
\langle\bar B(7) \, F(1) \rangle_{g^5} \, = \,
- \frac{g^5 N^3 (N^2-1)}{(4 \pi^2)^5} \left[
\frac{33}{\epsilon^2} + \frac{70}{\epsilon}+(-6 + 180 \, \zeta(3)) + O(\epsilon^1)
\right] \frac{1}{(x_{17}^2)^3} \, \left(x_{17}^2 \, \tilde \mu^2_x \right)^{5
\epsilon} \, .
\end{equation}

\newpage

\hskip - 0.5 cm
\begin{minipage}{\textwidth}
$G_1$ \hskip 2.74 cm $G_2$ \hskip 2.74 cm $G_3$ \hskip 2.74 cm $G_4$ \hskip 2.74 cm
$G_5$ \\
\includegraphics[width = \textwidth]{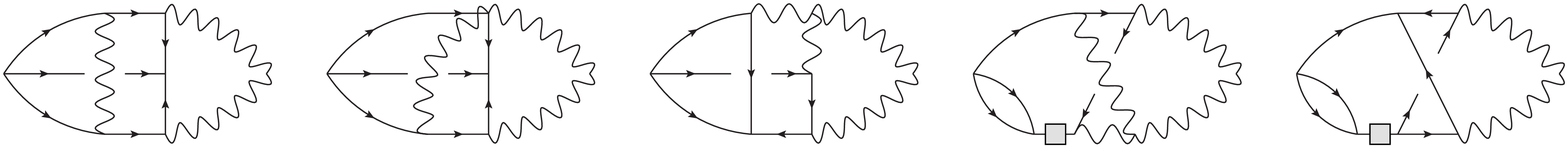} \\
$G_6$ \hskip 2.74 cm $G_7$ \hskip 2.74 cm $G_8$ \hskip 2.74 cm $G_9$ \hskip 2.74 cm
$G_{10}$ \\
\includegraphics[width = \textwidth]{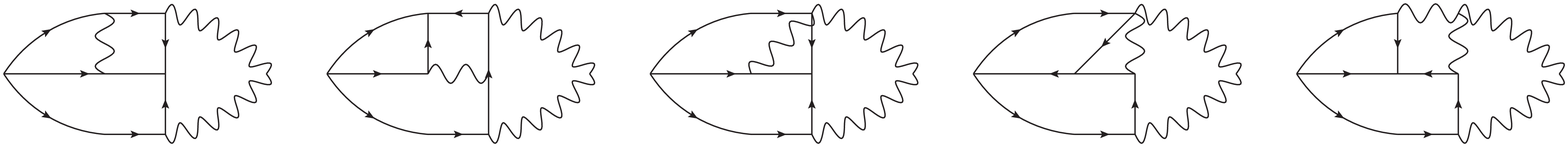} \\
$G_{11}$ \hskip 2.6 cm $G_{12}$ \hskip 2.6 cm $G_{13}$ \hskip 2.6 cm $G_{14}$
\hskip 2.6 cm $G_{15}$ \\
\includegraphics[width = 0.387 \textwidth]{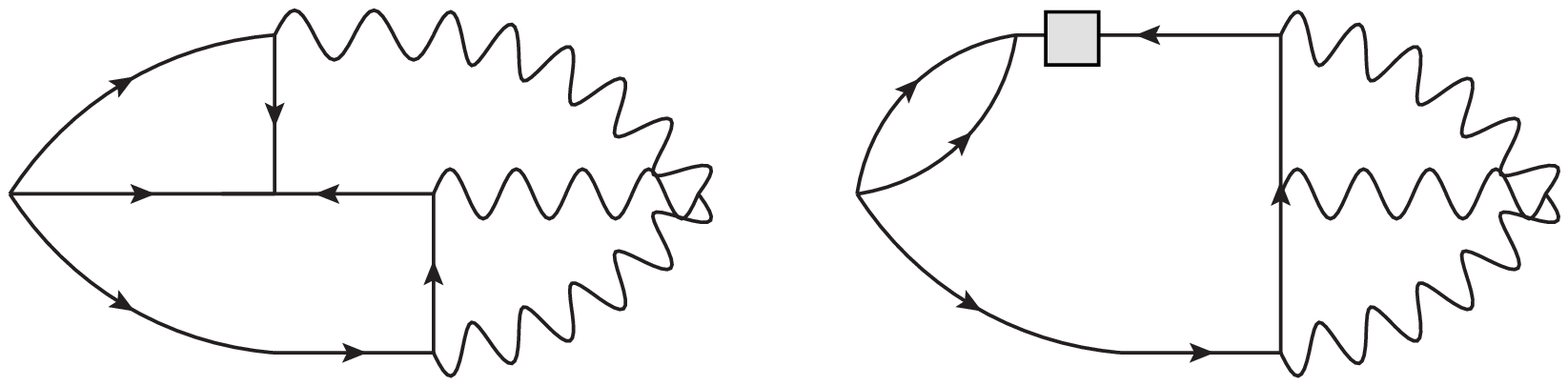} \hskip 0.019 \textwidth
\includegraphics[width = 0.591 \textwidth]{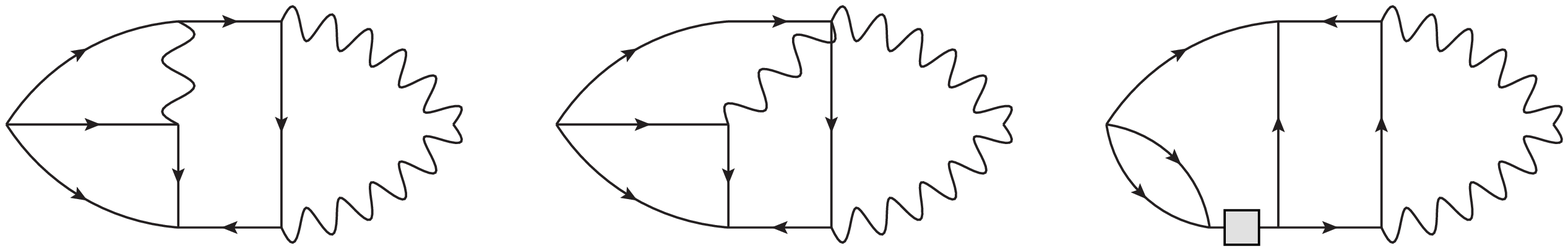} \\
$G_{16}$ \hskip 2.6 cm $G_{17}$ \hskip 2.6 cm $G_{18}$ \hskip 2.6 cm $G_{19}$
\hskip 2.6 cm $G_{20}$ \\
\includegraphics[width = \textwidth]{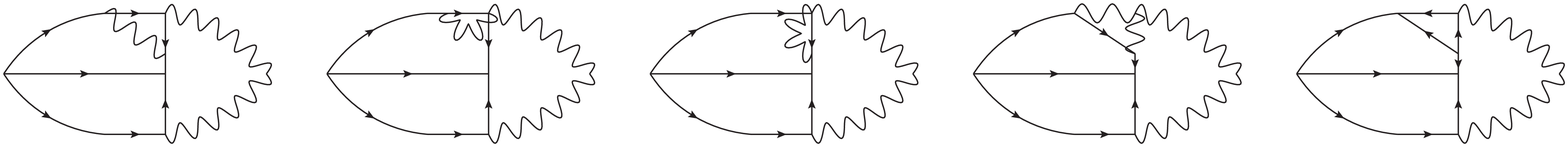} \\
$G_{21}$ \hskip 2.6 cm $G_{22}$ \hskip 2.6 cm $G_{23}$ \hskip 2.6 cm $G_{24}$
\hskip 2.6 cm $G_{25}$ \\
\includegraphics[width = \textwidth]{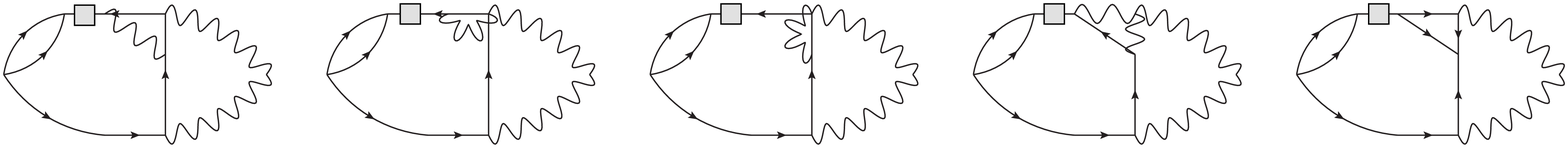} \\
$G_{26}$ \hskip 2.6 cm $G_{27}$ \hskip 2.6 cm $G_{28}$ \hskip 2.6 cm $G_{29}$
\hskip 2.6 cm $G_{30}$ \\
\includegraphics[width = \textwidth]{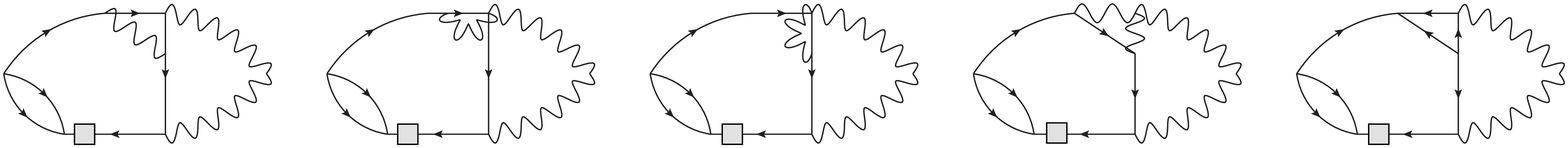} \\
$G_{31}$ \hskip 2.6 cm $G_{32}$ \hskip 2.6 cm $G_{33}$ \hskip 2.6 cm $G_{34}$ \\
\includegraphics[width = 0.387 \textwidth]{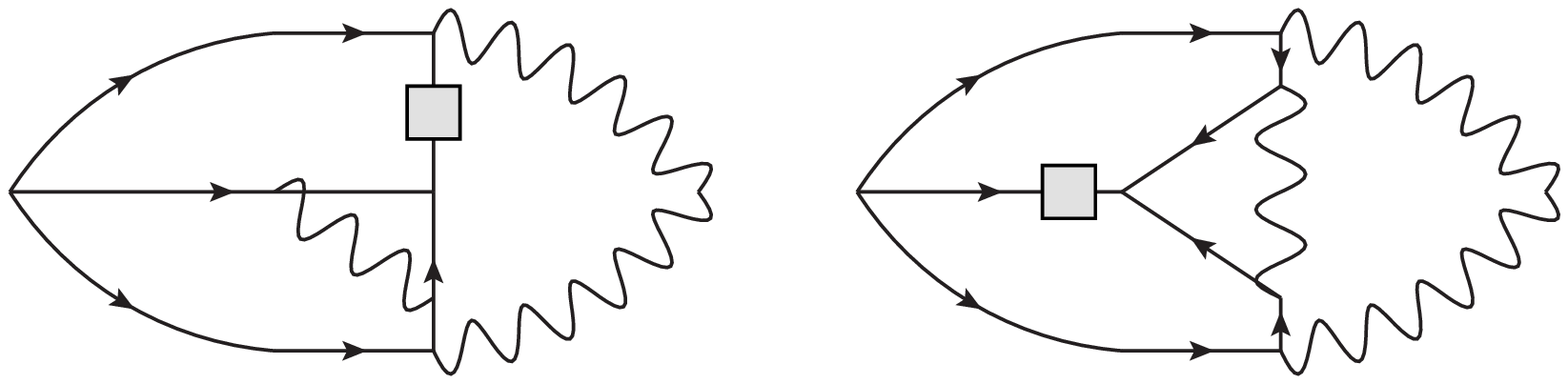} \hskip 0.016 \textwidth
\includegraphics[width = 0.387 \textwidth]{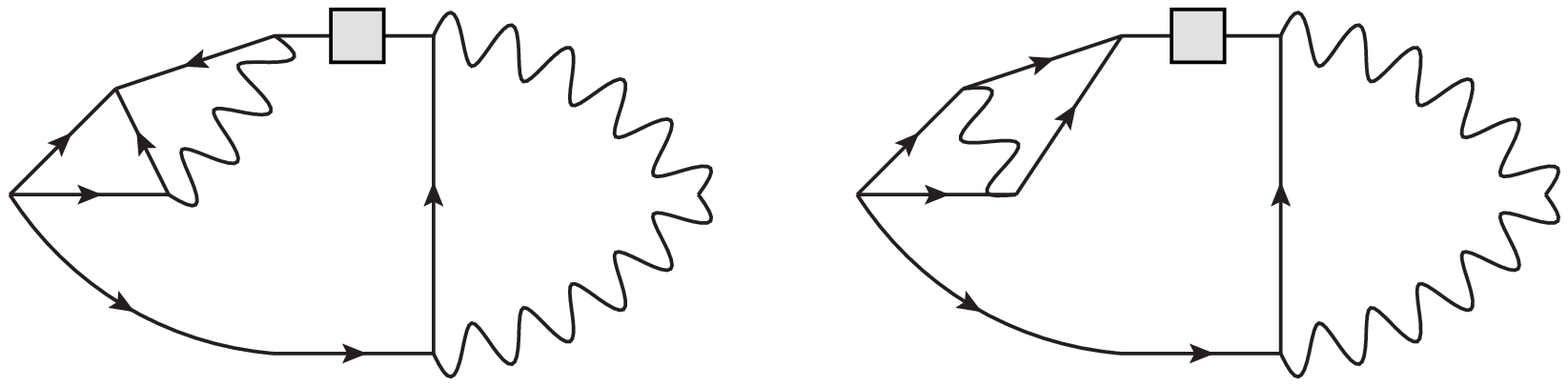} \\
$G_{35}$ \hskip 2.6 cm $G_{36}$ \hskip 2.6 cm $G_{37}$ \hskip 2.6 cm $G_{38}$ \\
\includegraphics[width = 0.796 \textwidth]{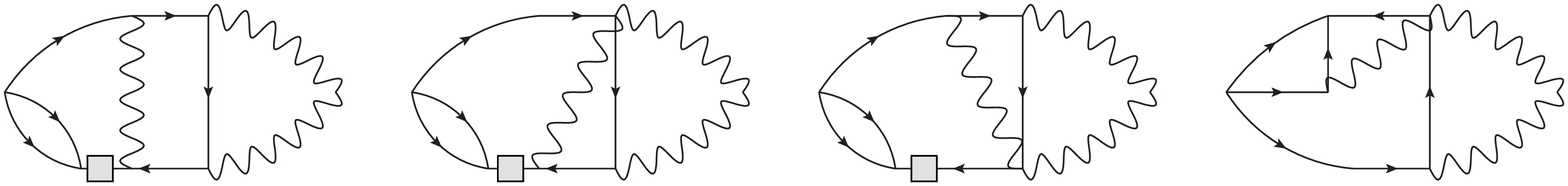} \\
$G_{39}$ \hskip 2.6 cm $G_{40}$ \hskip 2.6 cm $G_{41}$ \hskip 2.6 cm $G_{42}$
\hskip 2.6 cm $G_{43}$ \\
\includegraphics[width = \textwidth]{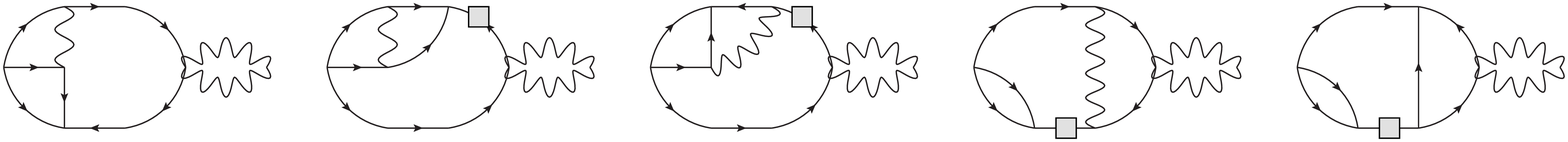} \\
$G_{44}$ \hskip 2.6 cm $G_{45}$ \hskip 2.6 cm $G_{46}$ \hskip 2.6 cm $G_{47}$
\hskip 2.6 cm $G_{48}$ \\
\includegraphics[width = \textwidth]{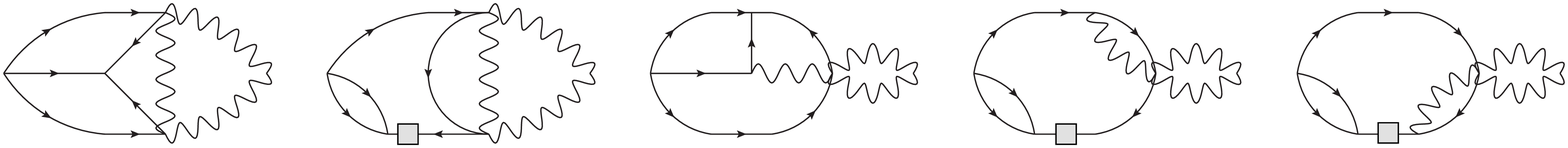}
\end{minipage}
\vskip -0.1 cm
\phantom{W} \hskip 5 cm \textbf{Figure 6}
\vskip 0.2 cm
Diagrams with vanishing colour factors (w.r.t. $SU(N)$ gauge group) or one-loop
propagator bubbles were omitted.
Graphs $G_{22}, \, G_{23}, \, G_{27}, \, G_{28}, \, G_{33} \ldots G_{48}$ do not need numerator
reduction by the Laporta ansatz, though the algorithm is needed here, too,
to determine some $\mathbf{T_1}$ configurations (see below) with one or two
non-integer exponents.

In (\ref{bigEq}) we give the momentum space result for each graph including all signs and
factors. The $\epsilon$ expansions in the table must be multiplied by a global factor
$- 12 \, g^5 N^3 (N^2-1) / (4 \pi)^8 * q^2 \, \left(q^2/\tilde \mu^2\right)^{-4 \epsilon}$.

\newpage

\phantom{werner}
\vskip - 1 cm
\begin{equation}
    \begin{array}{ccccccc}
    G_{1} & = & & \phm \frac{1}{4} \, \frac{1}{\epsilon^3} & + \left(
      \frac{101}{24}+\frac{1}{2} {\footnotesize \text{$\zeta(3)$}} \right) \, \frac{1}{\epsilon^2} & + \left(
      \frac{2077}{48}+\frac{1}{120}\, \pi^4+\frac{13}{4} \, {\footnotesize \text{$\zeta(3)$}} \right)\,
     \frac{1}{\epsilon} & + \, {\footnotesize \text{$O(\epsilon^0)$}}
      \\
    G_{2} & = & & -\frac{1}{8} \, \frac{1}{\epsilon^3} & - \left(
      \frac{113}{48}-\frac{1}{4} {\footnotesize \text{$\zeta(3)$}} \right) \, \frac{1}{\epsilon^2} &
      - \left(\frac{843}{32}-\frac{1}{240} \, \pi^4 -\frac{13}{8} \, {\footnotesize \text{$\zeta(3)$}} \right) \,
      \frac{1}{\epsilon} & + \, {\footnotesize \text{$O(\epsilon^0)$}}
      \\
    G_{3} & = & \phm \frac{1}{8} \, \frac{1}{\epsilon^4} & + \frac{83}{48} \,
      \frac{1}{\epsilon^3} & + \frac{483}{32} \, \frac{1}{\epsilon^2} & + \left( \frac{20507}{192}-\frac{35}{3}
       \, {\footnotesize \text{$\zeta(3)$}} + {\footnotesize \text{$5 \, \zeta (5)$}} \right) \, \frac{1}{\epsilon} &
      + \, {\footnotesize \text{$O(\epsilon^0)$}} \\
    G_{4} & = & -\frac{1}{8} \, \frac{1}{\epsilon^4} & -\frac{31}{16} \,
      \frac{1}{\epsilon^3} & -\frac{1765}{96} \, \frac{1}{\epsilon^2} & - \left( \frac{26881}{192} -
      \frac{73}{6} \, {\footnotesize \text{$\zeta(3)$}} \right) \, \frac{1}{\epsilon} & + \,
      {\footnotesize \text{$O(\epsilon^0)$}} \\
    G_{5} & = & & -\frac{2}{3} \, \frac{1}{\epsilon^3} &
      -\frac{37}{6} \, \frac{1}{\epsilon^2} & - \left(\frac{95}{4}-{\footnotesize \text{$3 \, \zeta(3)$}} \right) \,
      \frac{1}{\epsilon} & + \, {\footnotesize \text{$O(\epsilon^0)$}} \\
    G_{6} & = & & \phm \frac{1}{4} \, \frac{1}{\epsilon^3} & + \left(
      \frac{35}{8}-\frac{1}{2} {\footnotesize \text{$\zeta(3)$}} \right) \, \frac{1}{\epsilon^2} & + \left(\frac{739}{16}-
      \frac{1}{120} \, \pi^4-\frac{29}{4} \, {\footnotesize \text{$\zeta(3)$}} \right) \, \frac{1}{\epsilon} &
      + \, {\footnotesize \text{$O(\epsilon^0)$}} \\
    G_{7} & = & & -\frac{2}{3} \, \frac{1}{\epsilon^3} & - \left(
      \frac{21}{2}-{\footnotesize \text{$\zeta(3)$}} \right) \, \frac{1}{\epsilon^2} & - \left(\frac{401}{4}-
      \frac{1}{60} \, \pi^4 - \frac{21}{2} \, {\footnotesize \text{$\zeta(3)$}} \right) \, \frac{1}{\epsilon} &
      + \, {\footnotesize \text{$O(\epsilon^0)$}} \\
    G_{8} & = & &  & - \left(
      \frac{1}{2}-\frac{3}{2} {\footnotesize \text{$\zeta(3)$}} \right) \, \frac{1}{\epsilon^2} &
      - \left(\frac{37}{4} - \frac{1}{40} \, \pi^4 - \frac{67}{4} \, {\footnotesize \text{$\zeta(3)$}} \right) \,
      \frac{1}{\epsilon} & + \, {\footnotesize \text{$O(\epsilon^0)$}} \\
    G_{9} & = & & -\frac{1}{8} \, \frac{1}{\epsilon^3} &
      -\frac{113}{48} \, \frac{1}{\epsilon^2} & -\frac{843}{32} \, \frac{1}{\epsilon} & + \,
      {\footnotesize \text{$O(\epsilon^0)$}} \\
    G_{10} & = & \phm \frac{1}{8} \, \frac{1}{\epsilon^4} & + \frac{47}{24}
      \, \frac{1}{\epsilon^3} & + \frac{1789}{96} \, \frac{1}{\epsilon^2} & + \left( \frac{13543}{96}
      -\frac{179}{12} \, {\footnotesize \text{$\zeta(3)$}} \right) \, \frac{1}{\epsilon} &
      + \, {\footnotesize \text{$O(\epsilon^0)$}} \\
    G_{11} & = & & \phm \frac{3}{4} \, \frac{1}{\epsilon^3} &
      + \frac{93}{8} \, \frac{1}{\epsilon^2} & + \left( \frac{1765}{16}-{\footnotesize \text{$3 \, \zeta(3)$}} \right) \,
      \frac{1}{\epsilon} & + \, {\footnotesize \text{$O(\epsilon^0)$}} \\
    G_{12} & = & & -\frac{3}{4} \, \frac{1}{\epsilon^3} &
      -\frac{87}{8} \, \frac{1}{\epsilon^2} & -\left(\frac{1551}{16}-{\footnotesize \text{$6 \, \zeta(3)$}}+
        {\footnotesize \text{$5 \, \zeta (5)$}} \right)
      \, \frac{1}{\epsilon} & + \, {\footnotesize \text{$O(\epsilon^0)$}} \\
    G_{13} & = & & -\frac{1}{2} \, \frac{1}{\epsilon^3} &
      - \left(\frac{35}{4}+{\footnotesize \text{$\zeta(3)$}}\right) \, \frac{1}{\epsilon^2} & - \left(\frac{739}{8}+
      \frac{1}{60} \, \pi^4 + \frac{13}{2} \, {\footnotesize \text{$\zeta(3)$}} \right) \, \frac{1}{\epsilon} &
      + \, {\footnotesize \text{$O(\epsilon^0)$}} \\
    G_{14} & = & & \phm \frac{1}{8} \, \frac{1}{\epsilon^3} & + \left(
      \frac{59}{16}-\frac{3}{4} {\footnotesize \text{$\zeta(3)$}} \right) \, \frac{1}{\epsilon^2} & + \left(
      \frac{1579}{32}-\frac{1}{80} \, \pi^4-\frac{51}{8} \,  {\footnotesize \text{$\zeta(3)$}} \right) \,
      \frac{1}{\epsilon} & + \, {\footnotesize \text{$O(\epsilon^0)$}} \\
    G_{15} & = & & \phm {\footnotesize \text{3}} \, \frac{1}{\epsilon^3} & \phm \frac{53}{2} \,
      \frac{1}{\epsilon^2} & + \frac{405}{4} \, \frac{1}{\epsilon} & + \, {\footnotesize \text{$O(\epsilon^0)$}} \\
    G_{16} & = & & -\frac{2}{3} \, \frac{1}{\epsilon^3} &
      -\frac{37}{3} \, \frac{1}{\epsilon^2} & - \left( \frac{815}{6}- {\footnotesize \text{$6 \, \zeta(3)$}} \right) \,
      \frac{1}{\epsilon} & + \, {\footnotesize \text{$O(\epsilon^0)$}} \\
    G_{17} & = & & \phm \frac{5}{8} \, \frac{1}{\epsilon^3} & +
      \frac{183}{16} \, \frac{1}{\epsilon^2} & + \frac{4007}{32} \, \frac{1}{\epsilon} & + \,
      {\footnotesize \text{$O(\epsilon^0)$}} \\
    G_{18} & = & & \phm \frac{5}{8} \, \frac{1}{\epsilon^3} & +
      \frac{187}{16} \, \frac{1}{\epsilon^2} & + \frac{4115}{32} \, \frac{1}{\epsilon} &
      + \, {\footnotesize \text{$O(\epsilon^0)$}} \\
    G_{19} & = & & \phm \frac{1}{2} \, \frac{1}{\epsilon^3} & +
      \frac{47}{6} \, \frac{1}{\epsilon^2} & + \left(\frac{224}{3}- {\footnotesize \text{$3 \, \zeta(3)$}} \right) \,
      \frac{1}{\epsilon} & + \, {\footnotesize \text{$O(\epsilon^0)$}} \\
    G_{20} & = & & -\frac{4}{3} \, \frac{1}{\epsilon^3} &
      -\frac{68}{3} \, \frac{1}{\epsilon^2} & -{\footnotesize \text{233}} \, \frac{1}{\epsilon} &
      + \, {\footnotesize \text{$O(\epsilon^0)$}} \\
    G_{21} & = & & \phm \frac{3}{4} \, \frac{1}{\epsilon^3} & +
      \frac{299}{24} \, \frac{1}{\epsilon^2} & + \frac{1993}{16} \, \frac{1}{\epsilon} &
      + \, {\footnotesize \text{$O(\epsilon^0)$}} \\
    G_{22} & = & & -\frac{1}{4} \, \frac{1}{\epsilon^3} &
      -\frac{57}{8} \, \frac{1}{\epsilon^2} & -\frac{1453}{16} \, \frac{1}{\epsilon} &
      + \, {\footnotesize \text{$O(\epsilon^0)$}} \\
    G_{23} & = & & -\frac{1}{2} \, \frac{1}{\epsilon^3} &
      -\frac{39}{4} \, \frac{1}{\epsilon^2} & -\frac{871}{8} \, \frac{1}{\epsilon} & + \,
      {\footnotesize \text{$O(\epsilon^0)$}} \\
    G_{24} & = & -\frac{1}{8} \, \frac{1}{\epsilon^4} & -\frac{23}{12} \,
      \frac{1}{\epsilon^3} & -\frac{1787}{96} \, \frac{1}{\epsilon^2} & -\left(\frac{7013}{48}-
      \frac{91}{6} \, {\footnotesize \text{$\zeta(3)$}} \right) \, \frac{1}{\epsilon} &
      + \, {\footnotesize \text{$O(\epsilon^0)$}} \\
    G_{25} & = & & \phm \frac{1}{2} \, \frac{1}{\epsilon^3} & +
      \frac{145}{12} \, \frac{1}{\epsilon^2} & + \frac{1187}{8} \, \frac{1}{\epsilon} &
      + \, {\footnotesize \text{$O(\epsilon^0)$}} \\
    G_{26} & = & & \phm \frac{7}{12} \, \frac{1}{\epsilon^3} & +
      \frac{269}{24} \, \frac{1}{\epsilon^2} & + \left(\frac{5965}{48}- {\footnotesize \text{$6 \, \zeta(3)$}} \right) \,
      \frac{1}{\epsilon} & + \, {\footnotesize \text{$O(\epsilon^0)$}} \\
    G_{27} & = & & -\frac{1}{2} \, \frac{1}{\epsilon^3} &
      -\frac{37}{4} \, \frac{1}{\epsilon^2} & -\frac{817}{8} \, \frac{1}{\epsilon} &
      + \, {\footnotesize \text{$O(\epsilon^0)$}} \\
    G_{28} & = & & -\frac{1}{2} \, \frac{1}{\epsilon^3} &
      -\frac{39}{4} \, \frac{1}{\epsilon^2} & -\frac{871}{8} \, \frac{1}{\epsilon} &
      + \, {\footnotesize \text{$O(\epsilon^0)$}} \\
    G_{29} & = & & -\frac{5}{8} \, \frac{1}{\epsilon^3} &
      -\frac{19}{2} \, \frac{1}{\epsilon^2} & - \left( \frac{2853}{32}- {\footnotesize \text{$3 \, \zeta(3)$}} \right) \,
      \frac{1}{\epsilon} & + \, {\footnotesize \text{$O(\epsilon^0)$}} \\
    G_{30} & = & & \phm \frac{7}{6} \, \frac{1}{\epsilon^3} & +
      \frac{233}{12} \, \frac{1}{\epsilon^2} & + \frac{4729}{24} \, \frac{1}{\epsilon} &
      + \, {\footnotesize \text{$O(\epsilon^0)$}} \\
    G_{31} & = & & \phm \frac{1}{4} \, \frac{1}{\epsilon^3} & + \left(
      \frac{35}{8}-{\footnotesize \text{$\zeta(3)$}} \right) \, \frac{1}{\epsilon^2} & + \left( \frac{739}{16}-
      \frac{1}{60} \, \pi^4 -\frac{21}{2} \, {\footnotesize \text{$\zeta(3)$}} \right) \, \frac{1}{\epsilon} &
      + \, {\footnotesize \text{$O(\epsilon^0)$}} \\
    G_{32} & = & & \phm \frac{1}{4} \, \frac{1}{\epsilon^3} & +
      \frac{101}{24} \, \frac{1}{\epsilon^2} & + \frac{2077}{48} \, \frac{1}{\epsilon} &
      + \, {\footnotesize \text{$O(\epsilon^0)$}} \\
    G_{33} & = & & -\frac{1}{6} \, \frac{1}{\epsilon^3} &
      -\frac{59}{12} \, \frac{1}{\epsilon^2} & -\frac{1579}{24} \, \frac{1}{\epsilon} &
      + \, {\footnotesize \text{$O(\epsilon^0)$}} \\
    G_{34} & = & &  &  &
      -{\footnotesize \text{$\zeta(3)$}} \, \frac{1}{\epsilon} & + \, {\footnotesize \text{$O(\epsilon^0)$}} \\
    G_{35} & = & & -\frac{7}{6} \, \frac{1}{\epsilon^3} &
      -\frac{121}{12} \, \frac{1}{\epsilon^2} & -\frac{745}{24} \, \frac{1}{\epsilon} &
      + \, {\footnotesize \text{$O(\epsilon^0)$}} \\
    G_{36} & = & & \phm \frac{9}{8} \, \frac{1}{\epsilon^3} & +
      \frac{159}{16} \, \frac{1}{\epsilon^2} & + \frac{1215}{32} \, \frac{1}{\epsilon} &
      + \, {\footnotesize \text{$O(\epsilon^0)$}} \\
    G_{37} & = & & \phm \frac{7}{8} \, \frac{1}{\epsilon^3} & +
      \frac{121}{16} \, \frac{1}{\epsilon^2} & + \frac{745}{32} \, \frac{1}{\epsilon} &
      + \, {\footnotesize \text{$O(\epsilon^0)$}} \\
    G_{38} & = & & \phm \frac{1}{8} \, \frac{1}{\epsilon^3} & +
      \frac{59}{16} \, \frac{1}{\epsilon^2} & + \frac{1579}{32} \, \frac{1}{\epsilon} &
      + \, {\footnotesize \text{$O(\epsilon^0)$}} \\
    G_{39} & = & &  &  &
      - {\footnotesize \text{$16 \, \zeta(3)$}} \, \frac{1}{\epsilon} & + \, {\footnotesize \text{$O(\epsilon^0)$}} \\
    G_{40} & = & &  &  &
      - {\footnotesize \text{$8 \, \zeta(3)$}} \, \frac{1}{\epsilon} & + \, {\footnotesize \text{$O(\epsilon^0)$}} \\
    G_{41} & = & & -\frac{4}{3} \, \frac{1}{\epsilon^3} &
      -\frac{52}{3} \, \frac{1}{\epsilon^2} & -\frac{412}{3} \, \frac{1}{\epsilon} &
      + \, {\footnotesize \text{$O(\epsilon^0)$}} \\
    G_{42} & = & & \phm \frac{4}{3} \, \frac{1}{\epsilon^3} & +
      \frac{52}{3} \, \frac{1}{\epsilon^2} & + \frac{412}{3} \, \frac{1}{\epsilon} &
      + \, {\footnotesize \text{$O(\epsilon^0)$}} \\
    G_{43} & = & & -\frac{16}{3} \, \frac{1}{\epsilon^3} & - {\footnotesize \text{64}}
      \, \frac{1}{\epsilon^2} & -\frac{1408}{3} \, \frac{1}{\epsilon} & + \, {\footnotesize \text{$O(\epsilon^0)$}} \\
    G_{44} & = & &  & \phm \frac{1}{6} \,
     \frac{1}{\epsilon^2} & + \frac{13}{4} \, \frac{1}{\epsilon} & + \, {\footnotesize \text{$O(\epsilon^0)$}} \\
    G_{45} & = & & -\frac{5}{4} \, \frac{1}{\epsilon^3} &
      -\frac{111}{8} \, \frac{1}{\epsilon^2} & -\frac{1391}{16} \, \frac{1}{\epsilon} &
      + \, {\footnotesize \text{$O(\epsilon^0)$}} \\
    G_{46} & = & & \phm \frac{10}{9} \, \frac{1}{\epsilon^3} & +
      \frac{130}{9} \, \frac{1}{\epsilon^2} & + \frac{1030}{9} \, \frac{1}{\epsilon} &
      + \, {\footnotesize \text{$O(\epsilon^0)$}} \\
    G_{47} & = & & -\frac{10}{9} \, \frac{1}{\epsilon^3} &
      -\frac{130}{9} \, \frac{1}{\epsilon^2} & -\frac{1030}{9} \, \frac{1}{\epsilon} &
      + \, {\footnotesize \text{$O(\epsilon^0)$}} \\
    G_{48} & = & & -\frac{20}{9} \, \frac{1}{\epsilon^3} &
      -\frac{80}{3} \, \frac{1}{\epsilon^2} & -\frac{1760}{9} \, \frac{1}{\epsilon} &
      + \, {\footnotesize \text{$O(\epsilon^0)$}}
   \end{array} \label{bigEq}
\end{equation}

\section{$\langle\bar B B\rangle_{g^4}$}

At tree-level
\begin{equation}
\langle \bar B(2) \, B(1) \rangle_{g^0} \, = \, - \frac{2 N (N^2-1)}{(4 \pi^2)^3}
\frac{1}{(x_{12}^2)^3} \, \left(x_{12}^2 \, \tilde \mu_x^2\right)^{3 \epsilon} \,
(1 + O(\epsilon^3))
\end{equation}
where the $\epsilon^3 \zeta(3) + \ldots$ correction arises from
$\Gamma(1-\epsilon)$ in the numerator of the modified propagator in dimensional regularisation.
At order $g^2$ there are two graphs:
\begin{equation}
\langle \bar B(2) \, B(1) \rangle_{g^2} \, = \, 12 \, g^2 N^2 (N^2-1) \, \langle \bar \phi(2)
\phi(1) \rangle_{g^0} \, \left( H_1 \, + \, H_2 \right)
\end{equation}

\vskip 0.2 cm
\hskip -0.5 cm
\begin{minipage}{\textwidth}
$H_1$ \hskip 2.9 cm $H_2$ \\
\includegraphics[width = 0.4 \textwidth]{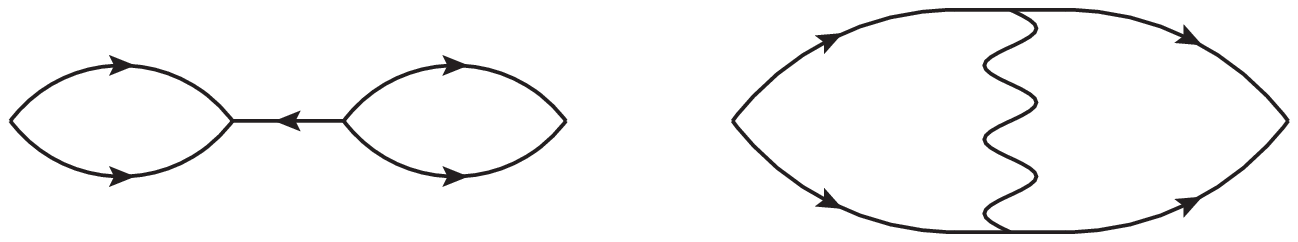}
\end{minipage}
\phantom{W} \hskip 2 cm \textbf{Figure 7}
\begin{eqnarray}
H_1 & = & - \frac{1}{(4 \pi)^4} \left[ \frac{1}{\epsilon^2} + \frac{4}{\epsilon} + 12 +
O(\epsilon) \right] \, \left(\frac{q^2}{\tilde \mu^2}\right)^{- 2 \epsilon} \, , \label{h1h2} \\
H_2 & = & - \frac{1}{(4 \pi)^4} \, \left[ \, 3 \, \zeta(3) + O(\epsilon) \, \right] \,
\left(\frac{q^2}{\tilde \mu^2}\right)^{- 2 \epsilon} \nonumber
\end{eqnarray}
whereby in configuration space
\begin{equation}
\langle \bar B(2) \, B(1) \rangle_{g^2} \, = \, - \frac{g^2 N}{4 \pi^2}
\langle \bar B(2) \, B(1) \rangle_{g^0} \, \left[ \frac{3}{\epsilon} + 3 + 9 \, \zeta(3) \,
\epsilon + O(\epsilon^2) \right] \, (x^2_{12} \, \tilde \mu_x^2)^\epsilon \, .
\end{equation}
The diagram $H_2$ is of topology $\mathbf{T_1}$ in the nomenclature of \cite{Mincer}.
Its numerator is just an outer box operator so that we can take over the result (\ref{t1res})
elaborated below. The leading term is of order $\epsilon^0$ in momentum space, which means
$O(\epsilon)$ in configuration space. It is therefore a contact term \cite{contact}.
Note that the contact contribution in (\ref{ffg2}) is exactly $H_2$ with a second
outer box operator, which does not touch upon the order in $\epsilon$
of the leading term.

On the other hand, w.r.t. the $O(g^4)$ part of $\langle \bar B \, B \rangle$ we are only
interested in the singular part in momentum space, or equivalently the singular and finite
pieces in configuration space. On comparing to a certain protected correlator we can spare most
of the work: Expanding in $N$, protected correlation functions contain several linear combinations
of graphs in which the $x$ space singular and finite terms must cancel.
One such sum of graphs is also present in $\langle \bar B \, B \rangle_{g^4}$, with identical
relative coefficients for the relevant graphs. Using the work of \cite{gamma3}:
\begin{equation}
\frac{\langle \bar \phi(2) \, \phi(1) \rangle_{g^0}}{12 \, g^4 N^3 (N^2-1)} \,
\langle \bar B(2) \, B(1) \rangle_{g^4}  \, = \, \langle \bar \phi(2) \,
\phi(1) \rangle_{g^0} \, \Sigma_1 \, + \, H_2^2 \, - \, \Sigma_2
\end{equation}
where $\Sigma_1$ contains the six four-loop graphs $H_3 \ldots H_8$
\vskip 0.5 cm
\hskip -0.5 cm
\begin{minipage}{\textwidth}
$H_3$ \hskip 2.18 cm $H_4$  \hskip 2.18 cm $H_5$ \hskip 2.18 cm $H_6$
\hskip 2.18 cm $H_7$ \hskip 2.18 cm $H_8$ \\
\includegraphics[width = \textwidth]{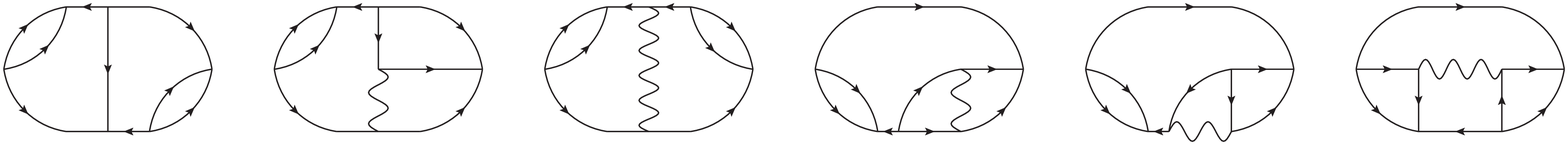}
\end{minipage}
\phantom{W} \hskip 5 cm \textbf{Figure 8}
\vskip 0.5 cm
\noindent and
\begin{equation}
\Sigma_2 \, = \, \left[ \frac{\langle \bar P(2) \, P(1) \,
\rangle_{g^4}}{12 \, g^4 (N^2 - 1)} \right]_{N^4} \, , \qquad P \, = \, \mathrm{Tr}(
\phi^1 \phi^1 \phi^2 \phi^2) + \frac{1}{2} \, \mathrm{Tr}(
\phi^1 \phi^2 \phi^1 \phi^2)
\end{equation}
is a sum of graphs in which only a contact contribution survives because $P$ is
an $SU(4)$ component of the half BPS operator $\mathrm{Tr}(\phi^1 \phi^1 \phi^1
\phi^1)$. Hence in $x$ space
\begin{equation}
\langle \bar B(2) \, B(1) \rangle_{g^4}  \, = \,
12 \, g^4 N^3 (N^2-1) \, \Sigma_1 \, + \, O(\epsilon) \, .
\end{equation}
The first five graphs in $\Sigma_1$ are simple to evaluate with the methods
developed for $\langle \bar B F \rangle_{g^5}$. For convenience
we used \emph{Mincer} for the three-loop ladder subgraph in $H_8$. Including
all signs and factors
\begin{equation}
   \begin{array}{cccccc}
    H_3 & = & \phm \phantom{\frac{1}{4}} \, \frac{1}{\epsilon^3} & +
      {\footnotesize \text{14}} \, \frac{1}{\epsilon^2} & + {\footnotesize \text{121}}
      \, \frac{1}{\epsilon} &
      + \, {\footnotesize \text{$O(\epsilon^0)$}} \\
    H_4 & = & \phm \frac{1}{4} \, \frac{1}{\epsilon^3} & + \frac{33}{8} \,
      \frac{1}{\epsilon^2} & + \left(\frac{665}{16}+ {\footnotesize \text{$6 \, \zeta(3)$}}
      \right) \, \frac{1}{\epsilon} &
      + \, {\footnotesize \text{$O(\epsilon^0)$}} \\
    H_5 & = & -\frac{1}{4} \, \frac{1}{\epsilon^3} & -\frac{33}{8} \, \frac{1}{\epsilon^2}
      & -\frac{657}{16} \, \frac{1}{\epsilon} &
      + \, {\footnotesize \text{$O(\epsilon^0)$}} \\
    H_6 & = & & & \frac{9}{2} {\footnotesize \text{$\, \zeta(3)$}} \, \frac{1}{\epsilon} &
      + \, {\footnotesize \text{$O(\epsilon^0)$}} \\
    H_7 & = & \phm \frac{3}{4} \, \frac{1}{\epsilon^3} & + \frac{87}{8} \,
      \frac{1}{\epsilon^2} & + \frac{1551}{16} \, \frac{1}{\epsilon} &
      + \, {\footnotesize \text{$O(\epsilon^0)$}} \\
    H_8 & = & -\frac{1}{4} \, \frac{1}{\epsilon^3} & -\frac{31}{8} \,
      \frac{1}{\epsilon^2} & -\frac{591}{16} \, \frac{1}{\epsilon} &
      + \, {\footnotesize \text{$O(\epsilon^0)$}}
   \end{array}
\end{equation}
to be multiplied by an overall $12 \, g^4 N^3 (N^2-1) / (4 \pi)^8 * q^2 \,
\left(q^2/\tilde \mu^2\right)^{-4 \epsilon}$. Summing up and translating to $x$ space
\begin{equation}
\langle \bar B(2) \, B(1) \rangle_{g^4} \, = \, \left(\frac{g^2 N}{4 \pi^2}\right)^2
\langle \bar B(2) \, B(1) \rangle_{g^0} \, \left[ \frac{9}{2} \, \frac{1}{\epsilon^2}
+\frac{45}{4} \, \frac{1}{\epsilon}  + \left( \frac{45}{4} + \frac{63}{2} \, \zeta(3)
\right) + O(\epsilon) \right] \, (x^2_{12} \, \tilde \mu_x^2)^{2 \epsilon} \, .
\end{equation}

\section{$\langle\bar {\cal O} {\cal O} \rangle_{g^4}$ and
$\langle\bar F F\rangle_{g^4}$}

Let us first consider the two-point function of the half BPS operator
\begin{equation}
{\cal O} \, = \, \mathrm{Tr}(\phi^1 \phi^1) \, . \nonumber
\end{equation}
The two leading orders are
\begin{equation}
\langle \bar {\cal O}(2) \, {\cal O}(1) \rangle \, = \,
2 (N^2-1) \left[ \langle \bar \phi(2) \phi(1)
\rangle_{g^0}^2 \, + \, 4 \, g^2 N \, H_2 + O(g^4) \right]
\end{equation}
while the $O(g^4)$ part receives contributions by the following graphs:
\vskip 0.4 cm
\hskip -0.5 cm
\begin{minipage}{\textwidth}
$I_1$ \hskip 2.36 cm $I_2$  \hskip 2.36 cm $I_3$ \hskip 2.36 cm $I_4$
\hskip 2.36 cm $I_5$ \hskip 2.36 cm $I_6$ \\
\includegraphics[width = \textwidth]{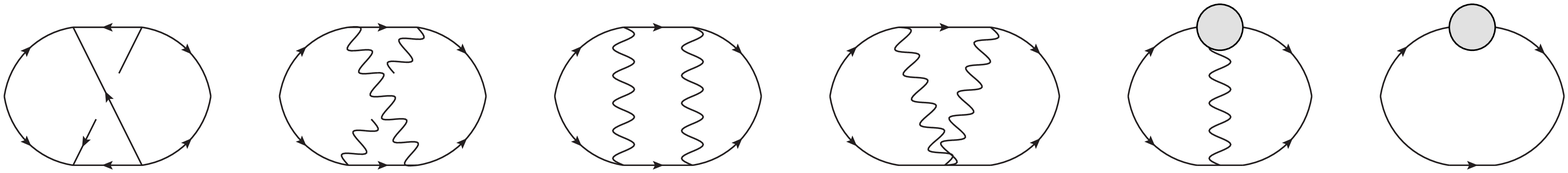}
\end{minipage}
\vskip -0.25 cm
\phantom{W} \hskip 5 cm \textbf{Figure 9}
\vskip 0.5 cm

The blob in $I_5$ means the one-loop correction to the cubic vertex
$g \int \mathrm{Tr}(\Phi^I [V,\bar \Phi_I])$ and the blob in $I_6$ denotes the two-loop correction to
the matter propagator.

\begin{equation}
   \begin{array}{ccccc}
    I_1 & = & \phm {\footnotesize \text{$\phantom{4} \, \zeta(3)$}} \, \frac{1}{\epsilon}
      & + \frac{\pi ^4}{60}+{\footnotesize \text{$6 \, \zeta(3)$}}
      & + \, O(\epsilon) \\
    I_2 & = & \phm {\footnotesize \text{$\phantom{4} \, \zeta(3)$}} \, \frac{1}{\epsilon}
      & + \frac{\pi ^4}{60}+{\footnotesize \text{$6 \, \zeta(3)$}}+{\footnotesize \text{$5 \,
      \zeta(5)$}} & + \, O(\epsilon) \\
    I_3 & = &  & \phm {\footnotesize \text{$20 \, \zeta(5)$}}
      & + \, O(\epsilon) \\
    I_4 & = &  & - {\footnotesize \text{$30 \, \zeta(5)$}}
      & + \, O(\epsilon) \\
    I_5 & = & - {\footnotesize \text{$4 \, \zeta(3)$}} \, \frac{1}{\epsilon} & - \frac{\pi^4}{15}-
      {\footnotesize \text{$24 \, \zeta(3)$}}+{\footnotesize \text{$30 \, \zeta(5)$}}
      & + \, O(\epsilon) \\
    I_6 & = & \phm {\footnotesize \text{$2 \, \zeta(3)$}} \, \frac{1}{\epsilon}
      & + \frac{\pi ^4}{30}+{\footnotesize \text{$12 \, \zeta(3)$}}
      & + \, O(\epsilon) \label{oog4}
   \end{array}
\end{equation}
The $\epsilon$ expansions in the table come with an overall factor $8 \, g^4 N^2 (N^2-1) / (4 \pi)^6 *
(q^2 / \tilde \mu^2)^{- 3 \epsilon}$. The sum of the lines in (\ref{oog4}) is $25 \, \zeta(5) + \,
O(\epsilon)$; hence only a contact term remains as required by protectedness.

We only give the sum of the propagator and vertex corrections because the loop order
here is not higher than three so that we could use the \emph{Mincer} programme to compute
the graphs. Moreover, the vanishing of the simple pole in the sum has been checked in
\cite{contact}.

Note that $I_1 \ldots I_3$ are only of order $\epsilon^{-1}$; in $x$ space they are therefore
individually finite, while $I_4$ is by itself a contact term. Likewise, the five graphs contributing
to $I_5$ are all separately finite in $x$ space. In ${\cal N} = 4$ SYM non-protected operators have
to be renormalised only due to diagrams with subdivergences coming from the outer ends; as
typical examples we mention the pictures of $H_1$ or $H_3, \, H_5$ above. For protected operators
these divergences
are absent. Due to the improved power counting in superspace, connected graphs tend to be
finite when they have sufficiently many outer legs. On the other hand, the individual graphs
contributing to
the (overall finite) two-loop matter blob have poles up to second order in momentum space
which the ``kinematical'' third loop in $I_6$ even promotes to $1/\epsilon^3$.

Collecting terms, in configuration space
\begin{eqnarray}
\langle \bar {\cal O}(2) \, {\cal O}(1) \rangle \, = \, \langle \bar
{\cal O}(2) \, {\cal O}(1) \rangle_{g^0} \, \bigg[1 & -
& \alpha\phantom{^2} \, \left(6 \, \zeta(3) \, \epsilon + O(\epsilon^2)\right) \,
(x_{12}^2 \tilde \mu_x^2)^\epsilon  \\
& + & \alpha^2 \, \left(\frac{75}{4} \, \zeta(5) \, \epsilon + O(\epsilon^2) \right) \,
(x_{12}^2 \tilde \mu_x^2)^{2 \epsilon} \, + \, O(\alpha^3) \bigg] \, ,
\qquad \alpha = \frac{g^2N}{4 \pi^2} \, .   \nonumber
\end{eqnarray}
Next, writing ${\cal O} \, = \, \mathrm{Tr}\left[ \left( e^{- g V} \phi^1 e^{g V} \right)
\left( e^{- g V} \phi^1 e^{g V} \right) \right]$ and using the equation of motion
\begin{equation}
D.D \, \left( e^{- g V} \phi^1 e^{g V} \right) \, = \, g \left[ \bar \phi_2, \bar \phi_3 \right] \label{eom}
\end{equation}
we derive
\begin{eqnarray}
\frac{1}{2} \, D^\alpha D_\alpha \, {\cal O} & = & \check F \, - \, 4 \, g \, \check B
\, , \nonumber \\ \check F & = & \mathrm{Tr} \left[ \left( D^\alpha
\left( e^{- g V} \phi^1 e^{g V} \right) \right) \, \left( D_\alpha \left(
e^{- g V} \phi^1 e^{g V} \right) \right) \right] \, , \nonumber \\ \check B &
= & \mathrm{Tr} \left[ \left( e^{- g V} \phi^1 e^{g V} \right) \, \left[ \bar
\phi_2, \, \bar \phi_3 \right] \right] \, . \nonumber
\end{eqnarray}
Recall that the $F$ involving two Yang-Mills fermions which we used before and the
representative $\check F$ are different $SU(4)$ components of the same
operator, likewise for $\check B$ and its ${\cal N}= 1$
chiral companion $B \, = \, \mathrm{Tr}(\phi^1 [\phi^2, \, \phi^3])$.

The Grassmann expansion of the correlator $\langle \bar {\cal O}(2) \,
{\cal O}(1) \rangle$
is given by the same exponential shift operator as for the superspace matter
propagator:
\begin{equation}
\langle \bar {\cal O}(2) \, {\cal O}(1) \rangle \, = \, e^{i \left(
(\theta_1 \partial_1 \bar \theta_1) + (\theta_2 \partial_1 \bar \theta_2) -
2 (\theta_1 \partial_1 \bar \theta_2) \right)} \,
\langle \bar {\cal O}(2) \, {\cal O}(1)
\rangle_{\theta_{1,2} = \bar \theta_{1,2} = 0}
\end{equation}
It follows that
\begin{equation}
\langle (\check F \, - \, 4 \, g \, \check B)(1) \,
(\bar {\check F} \, - \, 4 \, g \, \bar {\check B})(2) \rangle \, = \,
4 \, D.D|_1 \, \bar D . \bar D|_2 \, \langle \bar {\cal O}(2) \, {\cal O}
(1) \rangle \, = \, - 4 \, \square_1 \, \langle \bar {\cal O}(2) \,
{\cal O}(1) \rangle \, .
\end{equation}
In particular, for the $\theta = \bar \theta = 0$ component we should find
to leading order in $\epsilon$
\begin{eqnarray}
&& \langle (\check F \, - \, 4 \, g \, \check B)(1) \,
(\bar {\check F} \, - \, 4 \, g \, \bar {\check B})(2) \rangle \label{checkContact} \\ & = &
- \frac{64 (N^2-1)}{(4 \pi^2)^2 \left(x_{12}^2\right)^3}
\left(x_{12}^2 \, \tilde \mu_x^2 \right)^{2 \epsilon} \left[ (1 + O(\epsilon))
- \alpha \, \epsilon \, 6 \, \zeta(3) \left(x_{12}^2 \, \tilde \mu_x^2
\right)^{\epsilon}+ \alpha^2 \epsilon \frac{75}{4} \, \zeta(5) \left(x_{12}^2
\, \tilde \mu_x^2 \right)^{2 \epsilon}+ \ldots \right] \, . \nonumber
\end{eqnarray}
The tree-level part is obviously right for both $F$ and
$\check F$, whereas the $O(\alpha)$ contact term agrees with (\ref{ffg2})
because $F$ and $\check F$ have interchangeable two-point functions by
R-symmetry invariance.
We now wish to verify the $O(\alpha^2)$ contribution in the last equation by
a direct graph calculation. We prefer $\check F$ for this purpose because
$F = \lambda^\alpha \lambda_\alpha + \ldots$ would, of course, heavily involve
the Yang-Mills sector, while we have seen above that the D-algebra creates
quite some work already at $O(\alpha)$.

Upon expanding the exponentials in the definition in
(\ref{Fcheck})
\begin{eqnarray}
\check F & = & \mathrm{Tr} \Big( D^\alpha \phi^1 \, \label{expF}
D_\alpha \phi^1 \, - 2 \, g \, [ D^\alpha V, \, \phi^1 ] \, D_\alpha \phi^1 \, +
\\ && \phantom{\mathrm{Tr} \Big[} g^2 \, [ D^\alpha V, \, \phi^1 ] \,
[ D_\alpha V, \, \phi^1 ] \, + \,g^2 \, [ \, [ D^\alpha V, \, V], \phi^1 ] \,
D_\alpha \phi^1 \, + \,  O(g^3) \Big) \, . \nonumber
\end{eqnarray}
In the last formula $D^\alpha$ acts only on the field immediately after it.
In other words, in the given frame the chiral spinor derivative becomes
Yang-Mills covariantized:
\begin{equation}
\nabla^\alpha \, = \, D^\alpha \, - \, g \, [ D^\alpha V, \, \bullet \, ] \, +
\, \frac{g^2}{2} [ \, [ D^\alpha V, \, V ], \, \bullet \, ] \, + O(g^3)
\end{equation}
(On the other hand, the constraint $\bar D \, \phi^1 \, = \, 0$ is not
modified.)

The top graphs in $\langle \bar {\check F} \check F \rangle_{g^4, \theta = \bar \theta
= 0}$ are like in Figure 9 but with a partial spinor derivative on each outer
leg. These graphs have the $\epsilon$ expansions $\tilde I_1 \ldots \tilde I_6$
in the table below, which are visibly not overall derivatives of
$I_1 \ldots I_6$. Some new subdivergences have to be compensated by diagrams
arising from the higher terms in (\ref{expF}). Putting $\theta = \bar \theta
= 0$ at the outer points removes supergraphs with more than two outer $V$
fields. Further, most ways of placing
$D, \bar D$ on the fields at the outer points lead to vanishing results.
The remaining extra diagrams are given in Figure 10. We have put the derivatives
onto the graphs where their position could be ambiguous.
\vskip 0.2 cm
\hskip -0.5 cm
\begin{minipage}{\textwidth}
$\tilde I_{7\phantom{0}}$ \hskip 2.74 cm $\tilde I_{8\phantom{0}}$
\hskip 2.74 cm $\tilde I_{9\phantom{0}}$
\hskip 2.74 cm $\tilde I_{10}$ \hskip 2.74 cm $\tilde I_{11}$ \\
\includegraphics[width = \textwidth]{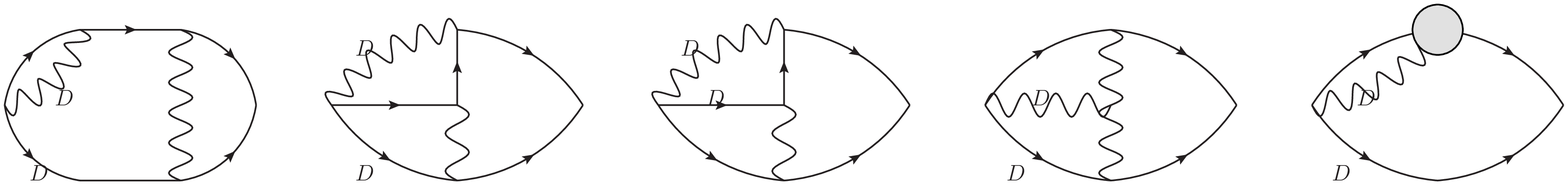} \\
\vskip -0.3 cm
$\tilde I_{12}$ \hskip 2.74 cm $\tilde I_{13}$ \hskip 2.74 cm $\tilde I_{14}$
\hskip 2.74 cm $\tilde I_{15}$ \hskip 2.74 cm $\tilde I_{16}$ \\
\includegraphics[width = 1.018 \textwidth]{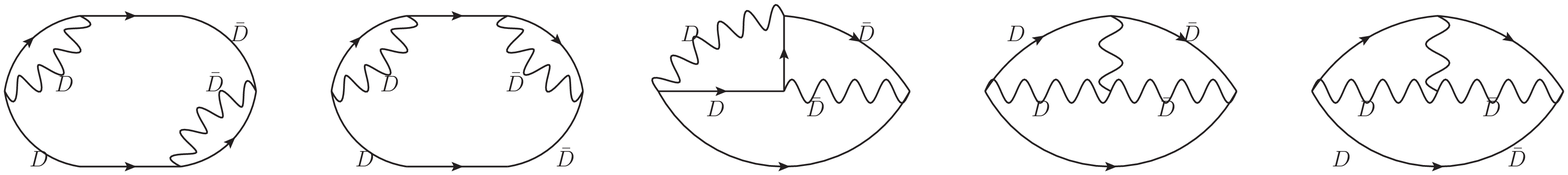} \\
\vskip -0.3 cm
$\tilde I_{17}$ \hskip 2.74 cm $\tilde I_{18}$ \hskip 2.74 cm $\tilde I_{19}$ \\
\includegraphics[width = 0.588 \textwidth]{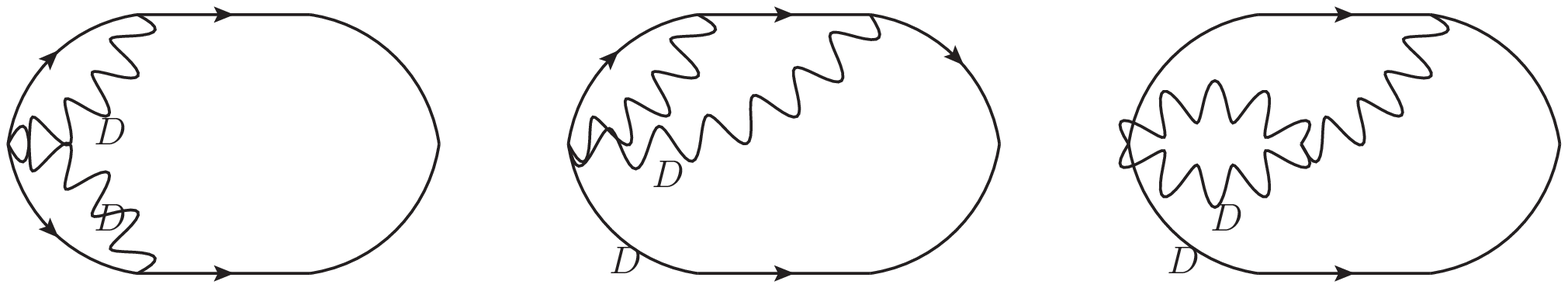} \\
\end{minipage}
\vskip - 0.7 cm
\phantom{W} \hskip 5 cm \textbf{Figure 10} \\
\begin{equation}
   \begin{array}{ccccccc}
    \tilde I_{1} & = & & -\frac{1}{6} \, \frac{1}{\epsilon^2} &
      -\left( \frac{4}{3}-\frac{1}{2} {\footnotesize \text{$\zeta(3)$}}\right)
      \, \frac{1}{\epsilon}
      & -\frac{19}{3}+\frac{\pi^4}{120}+{\footnotesize \text{$3 \, \zeta(3)$}}
      & + \, O(\epsilon) \\
    \tilde I_{2} & = & & -\frac{1}{6} \, \frac{1}{\epsilon^2} &
      -\left({\footnotesize \text{2}}-\frac{3}{2} {\footnotesize
      \text{$\zeta(3)$}} \right) \,
      \frac{1}{\epsilon} & -{\footnotesize \text{15}} +\frac{\pi^4}{40}+
      {\footnotesize \text{$7 \, \zeta(3)$}} +\frac{5}{2}
      {\footnotesize \text{$\zeta(5)$}} & + \, O(\epsilon) \\
    \tilde I_{3} & = & \phm \frac{2}{3} \, \frac{1}{\epsilon^3} & +
      \frac{16}{3} \, \frac{1}{\epsilon^2}
      & + \left( \frac{88}{3}- {\footnotesize \text{$12 \, \zeta(3)$}}\right)
      \, \frac{1}{\epsilon} &
      + \frac{416}{3}-\frac{\pi^4}{5}-\frac{196}{3}
      {\footnotesize \text{$\zeta(3)$}} + {\footnotesize \text{$10 \,
      \zeta(5)$}} & + \, O(\epsilon) \\
    \tilde I_{4} & = & \phm \frac{1}{2} \, \frac{1}{\epsilon^3} & +
      {\footnotesize \text{5}} \, \frac{1}{\epsilon^2} &
      + \frac{131}{4} \, \frac{1}{\epsilon} &
      + \frac{711}{4}-\frac{41}{2} {\footnotesize \text{$\zeta(3)$}}-
      {\footnotesize \text{$15 \, \zeta(5)$}} & + \, O(\epsilon) \\
    \tilde I_{5} & = & -\frac{1}{3} \, \frac{1}{\epsilon^3} & -\frac{10}{3}
      \, \frac{1}{\epsilon^2} & -\left(\frac{87}{4}+ {\footnotesize
      \text{$2 \, \zeta(3)$}} \right) \,
      \frac{1}{\epsilon} &
      -\frac{1409}{12}-\frac{\pi^4}{30}-\frac{10}{3} {\footnotesize
      \text{$\zeta(3)$}} + {\footnotesize \text{$15 \, \zeta(5)$}}
      & + \, O(\epsilon) \\
    \tilde I_{6} & = & &  & + {\footnotesize \text{$\zeta(3)$}}
      \, \frac{1}{\epsilon} & + \frac{\pi^4}{60}+{\footnotesize \text{$7 \,
      \zeta(3)$}} & + \, O(\epsilon) \\
    \tilde I_{7} & = & - {\footnotesize \text{2}} \, \frac{1}{\epsilon^3} & -
      {\footnotesize \text{18}} \, \frac{1}{\epsilon^2} &
      -\left(\frac{328}{3}- {\footnotesize \text{$12 \, \zeta(3)$}} \right) \,
      \frac{1}{\epsilon} & -\frac{1684}{3}+\frac{\pi^4}{5}+ {\footnotesize
      \text{$110 \, \zeta(3)$}} & + \, O(\epsilon) \\
    \tilde I_{8} & = & & \phm \frac{1}{6} \, \frac{1}{\epsilon^2} & + \left(
      {\footnotesize \text{2}} -
      {\footnotesize \text{$\zeta(3)$}} \right) \, \frac{1}{\epsilon} & +
      {\footnotesize \text{15}} -\frac{\pi^4}{60}-
      {\footnotesize \text{$3 \, \zeta(3)$}} & + \, O(\epsilon) \\
    \tilde I_{9} & = & -\frac{1}{6} \, \frac{1}{\epsilon^3} & -\frac{3}{2}
      \, \frac{1}{\epsilon^2} & -{\footnotesize \text{9}} \,
      \frac{1}{\epsilon} & -\frac{136}{3}+\frac{17}{6}
      {\footnotesize \text{$\zeta(3)$}} &
      + \, O(\epsilon) \\
    \tilde I_{10} & = & -\frac{1}{4} \, \frac{1}{\epsilon^3} & -\frac{5}{2}
      \, \frac{1}{\epsilon^2} & -\left(\frac{49}{3}+{\footnotesize
      \text{$\zeta(3)$}}\right) \,
      \frac{1}{\epsilon} &
      -\frac{265}{3}-\frac{\pi^4}{60}+\frac{13}{4} {\footnotesize
      \text{$\zeta(3)$}} &
      + \, O(\epsilon) \\
    \tilde I_{11} & = & \phm \frac{1}{6} \, \frac{1}{\epsilon^3} & + \frac{5}{3}
      \, \frac{1}{\epsilon^2} & + \left({\footnotesize \text{11}}+\frac{1}{2}
      {\footnotesize \text{$\zeta(3)$}} \right) \, \frac{1}{\epsilon} & +
      \frac{181}{3}+\frac{\pi
      ^4}{120}-\frac{4}{3} {\footnotesize \text{$\zeta(3)$}} & + \,
      O(\epsilon) \\
    \tilde I_{12} & = & \phm \frac{1}{3} \, \frac{1}{\epsilon^3} & +
      {\footnotesize \text{3}} \, \frac{1}{\epsilon^2} &
      + \frac{55}{3} \, \frac{1}{\epsilon} & + {\footnotesize \text{95}} -
      \frac{29}{3}  {\footnotesize \text{$\zeta(3)$}} & + \, O(\epsilon) \\
    \tilde I_{13} & = & \phm \frac{1}{3} \, \frac{1}{\epsilon^3} & +
      {\footnotesize \text{3}} \, \frac{1}{\epsilon^2} & + {\footnotesize
      \text{18}} \, \frac{1}{\epsilon} & + \frac{272}{3}-\frac{23}{3}
      {\footnotesize \text{$\zeta(3)$}}
      & + \, O(\epsilon) \\
    \tilde I_{14} & = & & & \frac{1}{6}
      \, \frac{1}{\epsilon} & + \frac{13}{6}-{\footnotesize \text{$\zeta(3)$}} &
      + \, O(\epsilon) \\
    \tilde I_{15} & = & & & \frac{1}{6}
      \, \frac{1}{\epsilon} & + \frac{13}{6} & + \, O(\epsilon) \\
    \tilde I_{16} & = & & & \frac{1}{2} {\footnotesize \text{$\zeta(3)$}} \,
      \frac{1}{\epsilon} &
      + \frac{\pi^4}{120}+\frac{7}{2} {\footnotesize \text{$\zeta(3)$}} &
      + \, O(\epsilon) \\
    \tilde I_{17} & = & \phm \frac{1}{2} \, \frac{1}{\epsilon^3} & +
      \frac{9}{2} \, \frac{1}{\epsilon^2}
      & + \frac{55}{2} \, \frac{1}{\epsilon} & + \frac{285}{2}-
      \frac{29}{2} {\footnotesize \text{$\zeta(3)$}} & + \, O(\epsilon) \\
    \tilde I_{18} & = & \phm \frac{1}{12} \, \frac{1}{\epsilon^3} & +
      \frac{5}{6} \, \frac{1}{\epsilon^2} & + \frac{11}{2} \,
      \frac{1}{\epsilon} & + \frac{181}{6}-\frac{29}{12} {\footnotesize
      \text{$\zeta(3)$}}
      & + \, O(\epsilon) \\
    \tilde I_{19} & = & \phm \frac{1}{6} \, \frac{1}{\epsilon^3} & +
      \frac{5}{3} \, \frac{1}{\epsilon^2}
      & + {\footnotesize \text{11}} \, \frac{1}{\epsilon} &
      + \frac{181}{3}-\frac{29}{6}  {\footnotesize \text{$\zeta(3)$}} &
      + \, O(\epsilon)
  \end{array}
\end{equation}
In the table we have omitted the overall factor
$F_0 \, = \, 64 \, g^4 N^2 (N^2-1)/(4 \pi)^6 * q^2 \left(q^2/\tilde \mu^2\right)^{-3\epsilon}$.
The sum of $\tilde I_1 \ldots \tilde I_{19}$ yields the first line in the next equation.
\begin{eqnarray}
\langle \bar F \, F \rangle_{g^4} & = & F_0 \left[ - \frac{1}{2} \frac{1}{\epsilon^2} -
\frac{4}{\epsilon} - 19 + 3 \, \zeta(3) + \frac{25}{2} \, \zeta(5) + O(\epsilon) \right] \, , \\
- 8 \, g \, \langle \bar F \, B \rangle_{g^3} & = & F_0 \left[ + \frac{3}{2} \frac{1}{\epsilon^2} +
\frac{14}{\epsilon} + 83 + O(\epsilon) \right] \, , \nonumber \\
16 \, g^2 \, \langle \bar B \, B \rangle_{g^2} & = & F_0 \left[ - \frac{1}{\epsilon^2} -
\frac{10}{\epsilon} - 64  - 3 \, \zeta(3) + O(\epsilon) \right] \, . \nonumber
\end{eqnarray}
Upon adding up
\begin{equation}
\langle (\bar F - 4 g \bar B) (F - 4 g B) \rangle_{g^4} \, = \, F_0 \, \frac{25}{2} \zeta(5) \, + \,
O(\epsilon)
\end{equation}
which reproduces the $O(\alpha^2)$ term in (\ref{checkContact}) after Fourier transform.

In conclusion, at the given orders the operator $F - 4 \, g \, B$ does not need
renormalisation. Its anomalous
dimension is zero at one and two loops. At loop level the two-point function does pick up
contact contributions; remarkably the $O(\alpha)$ contact term has normalisation
proportional to $\zeta(3)$ while the $O(\alpha^2)$ term comes with
$\zeta(5)$. The
equation of motion relating the operator $O = \mathrm{Tr}(\phi^1 \phi^1)$ to
$F - 4 \, g \, B$ is apparently purely classical: To leading order in
$\epsilon$ the normalisation of the contact part is compatible with superspace
differentiation.

\section{Integration by parts (IBP) in dimensional regularisation}

We follow the last reference in \cite{Mincer} in exposing the fundamental idea.
Suppose that any given diagram has a triangle subgraph like the first picture in Figure 11.
\vskip 0.2 cm
\hskip - 0.5 cm
\begin{minipage}{\textwidth}
\includegraphics[width = \textwidth]{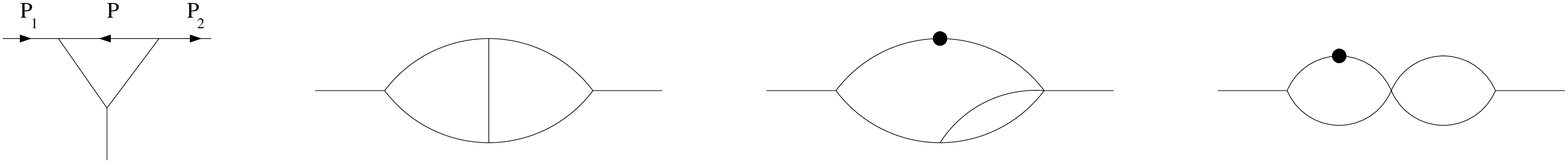}
\end{minipage}
\vskip -0.2 cm \phantom{W} \hskip 5 cm \textbf{Figure 11} \vskip
0.5 cm \noindent The integral associated to the subdiagram is
\begin{equation}
I(\alpha_0,\beta_1,\beta_2,\alpha_1,\alpha_2) \, = \, \int d^Dp \, \frac{1}
{\left(p^2\right)^{\alpha_0} \, \left((p+p_1)^2\right)^{\beta_1} \,
 \left(p_1^2\right)^{\alpha_1} \, \left((p+p_2)^2\right)^{\beta_2} \,
 \left(p_2^2\right)^{\alpha_2}} \, .
\end{equation}
Discarding boundary terms
\begin{equation}
\int d^Dp \; \partial_{p_\mu} \frac{p_\mu}
{\left(p^2\right)^{\alpha_0} \, \left((p+p_1)^2\right)^{\beta_1} \,
 \left(p_1^2\right)^{\alpha_1} \, \left((p+p_2)^2\right)^{\beta_2} \,
 \left(p_2^2\right)^{\alpha_2}} \, = \, 0
\end{equation}
because the integrand is a total derivative. By working out the differentiation:
\begin{eqnarray}
 && I(\alpha_0,\beta_1,\beta_2,\alpha_1,\alpha_2) \,
(D - 2 \, \alpha_0 - \beta_1 - \beta_2) \, = \, \label{triagRule} \\
&& \qquad \beta_1 \left( I(\alpha_0-1,\beta_1+1,\beta_2,\alpha_1,\alpha_2) -
I(\alpha_0,\beta_1+1,\beta_2,\alpha_1-1,\alpha_2) \right) + \nonumber \\
&& \qquad \beta_2 \left( I(\alpha_0-1,\beta_1,\beta_2+1,\alpha_1,\alpha_2) -
I(\alpha_0,\beta_1,\beta_2+1,\alpha_1,\alpha_2-1) \right) \nonumber
\end{eqnarray}
The generalisation of the formula for polynomial numerators
$p^{\nu_1} \ldots p^{\nu_n}$ is straightforward. In every term on
the r.h.s. of (\ref{triagRule}) the exponent of one of the
propagators of the triangle's top line is diminished by one. To
illustrate the use of the formula we compute the integral
\textbf{T}$_1$ (without numerator) given in the second picture of
Figure 11. Using (\ref{triagRule}) on the left triangle in the
diagram produces an overall factor $1/\epsilon$ times the third
minus the fourth picture. Our convention is that a line without
any extra symbol has exponent equal to one. The triangle rule can
remove such a line, but in exchange it augments the weight of one
of the $\beta$ lines by one. This is customarily denoted by a dot.
The resulting new diagrams are both elementary: One is a
convolution, the other the product of two one-loop integrals. The
elementary one-loop building block is
\begin{equation}
\int \frac{d^Dp}{(2 \, \pi)^D} \frac{1}{\left(p^2\right)^{\alpha} \,
\left((q-p)^2\right)^{\beta}} \, = \, \frac{1}{(4 \pi)^2} \, G(\alpha,\beta) \,
\left( q^2 \right)^{D/2 - \alpha - \beta} \, , \qquad D \, = \, 4 - 2 \,
\epsilon \label{int1l}
\end{equation}
where
\begin{equation}
G(\alpha,\beta) \, = \, (4 \pi)^\epsilon \, R(\alpha) \, R(\beta) \, R(D -
\alpha - \beta) \, , \qquad R(\alpha) \, = \, \frac{\Gamma(D/2 - \alpha)}
{\Gamma(\alpha)} \, .
\end{equation}
For the generalisation to integrals with numerator polynomials we refer to
\cite{Mincer}. It follows
\begin{equation}
\mathbf{T_1} \, = \, \frac{1}{\epsilon} \, G(1,1) \left( G(2,1+\epsilon) -
G(2,1) \right) \, = \, \frac{1}{(4 \, \pi)^4} \left( 6 \, \zeta(3) + \left(
\frac{\pi^4}{10} + 12 \, \zeta(3) \right) \epsilon + \ldots \right)
\frac{1}{q^2} \left(\frac{q^2}{\tilde \mu^2}\right)^{- 2 \epsilon} \label{t1res}
\end{equation}
in the $\overline{MS}$ convention explained above. Almost all
three-loop ``p-integrals'' (propagator type) can be calculated by
this trick \cite{Mincer}: By way of example, to solve the \textbf{BU} topology
displayed in Figure 3 one may start with one of the visible
triangles and then iterate the procedure on the resulting
\textbf{T}$_1$ (sub)diagrams.

Let us now turn to the four-loop topology \textbf{A} of our
$\langle  \bar B F \rangle_{g^5}$ problem. We can use the rule of
the triangle once: \vskip 0.5 cm \hskip - 0.52 cm
\begin{minipage}{\textwidth}
\includegraphics[width = 0.8 \textwidth]{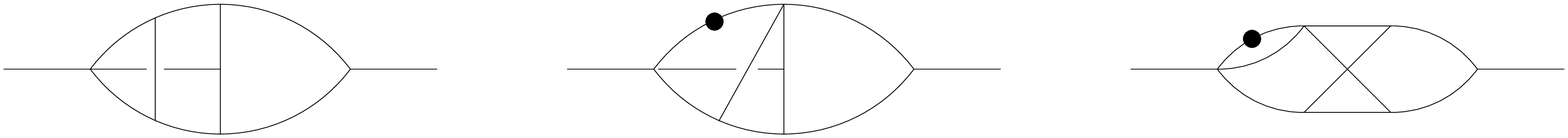}
\end{minipage}
\vskip -0.2 cm \phantom{W} \hskip 5 cm \textbf{Figure 12} \vskip
0.5 cm The two new topologies contain no further triangle.
Nonetheless, the IBP technique can be pushed further by the
Laporta ansatz \cite{Laporta}: The integral in the middle of
Figure 12 now has propagators with power one for, say, $p_1$ to
$p_8$ and of power two for $p_9$. The most complicated numerators
we encounter have three non-trivial dot products. We write the
complete set of identities
\begin{equation}
\int d^Dp_1 \ldots d^Dp_4 \; \partial_{p_i^\mu} \frac{p_j^\mu \, num }
{p_1^2 \, \ldots \, p_8^2 \, \left(p_9^2\right)^2} \, = \, 0 \label{laporta}
\end{equation}
where $i \in \{1 \ldots 4\}$ and $p_j$ can be one of the loop
momenta or the outer momentum $q$. Between the four loop momenta
and $q$ one can form fifteen independent square invariants. Let
the first ten be the squares of all the interior momenta and of
$q$, then we have to choose five further mixed dot products. Mixed
products containing $q$ bring an advantage. We write the 20
identities (\ref{laporta}) for any such numerator $num$ built from
three square invariants that does not by itself cancel a line in
the denominator (thus it can contain the mixed dot products, $q^2$
or one power of $p_9^2$). Next the differentiation is worked out
just as in the case of the triangle rule. The result is a large
homogeneous linear system for a basis of integrals, which one may
reduce by Gaussian eliminination. The matrix is initially very sparse and
the elimination has the surprising property that the relative
order in $\epsilon$ between the terms in the same line remains
relatively stable throughout the steps of the algorithm.

Due to memory limitations (1 GB on a power PC and later a
Xeon, of which we needed only about one half) we organised the task in
a recursive way: In the first step, except for the integral with a sought
numerator, all nine-propagator structures are eliminated. We obtain a
linear equation relating it to cases with eight or
less propagators (the cancellation of a line works
as in the case of the triangle rule). The set of eight-propagator
configurations is given in Figure 13. The differentiation in
(\ref{laporta}) produces a second dot, which we did not indicate on the
pictures because it can be placed on any of the interior lines. By dimensionality
the eight-propagator integrals still have numerators with
maximally three kinimatic invariants, or less if there is a
further cancellation with a denominator term. \vskip 0.5 cm \hskip
-0.52 cm
\begin{minipage}{\textwidth}
$M_{35}$ \hskip 3.87 cm $M_{36}$ \\
\includegraphics[width = 0.8 \textwidth]{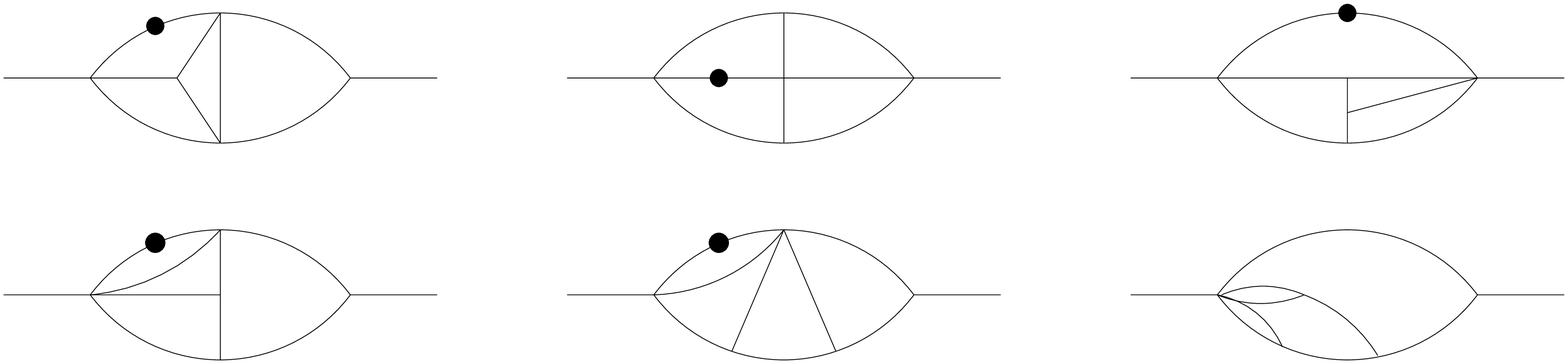}
\end{minipage}
\vskip - 0.2 cm \phantom{W} \hskip 5 cm \textbf{Figure 13} \vskip
0.5 cm
On this level, $M_{35}$ and $M_{36}$ once again need numerator
reduction by the Laporta ansatz. There is a new feature arising
here: Apart from the desired numerator term one can eliminate all
eight-propagator cases but one. The conventional choice for the
remaining integral is $M_{35}$ or $M_{36}$, respectively, with trivial
numerator $q^2$ and all exponents equal to one. These integrals
are called ``masters''. A priori they have to be calculated by
independent means. The attempt to eliminate a master usually
trivializes the system of equations.

The other integrals in Figure 13 and the seven-propagator
configurations found by cancelling a line from $M_{35}, \, M_{36}$
either have triangles or they are trivial like the last picture in
Figure 13. However, according to equation (\ref{int1l}) a one-loop
subintegral leaves behind a propagator with non-integer dimension;
in Figure 14 we marked this by a cross. If this affects one of the
$\alpha$ lines of a triangle subgraph, equation (\ref{triagRule})
ceases to be helpful. As a consequence, in the last step a variety
of \textbf{T}$_1$ cases with modified propagators again have to be
attacked by the Laporta idea. One can eventually backsubsitute
starting from a general result for \textbf{T}$_1(1 + a_1 \,
\epsilon, \, 1 + a_2 \, \epsilon,\, 1 + a_3 \, \epsilon,\, 1 + a_4
\, \epsilon,\, 1 + a_5 \, \epsilon)$ due to \cite{broadhurst} and
explicit results for the masters $M_{35}, \, M_{36}$ \cite{masters}.

The integral in the third picture in Figure 12 can be dealt with
in the same manner: The bubble integral leads to the
three-loop \textbf{NO} topology with a cross on one of the outer
lines. Once again, the Laporta idea is needed to further reduce
this. Foreseeably, there is a master integral for which we choose
$\mathrm{NO}(\epsilon,1,1,1,1,1,1,1)$. In Figure 14 we list the
more trivial seven-propagator configurations, of which only the first
case needs Laporta reduction.
\vskip 0.5 cm
\hskip - 0.5 cm
\begin{minipage}{\textwidth}
\includegraphics[width = \textwidth]{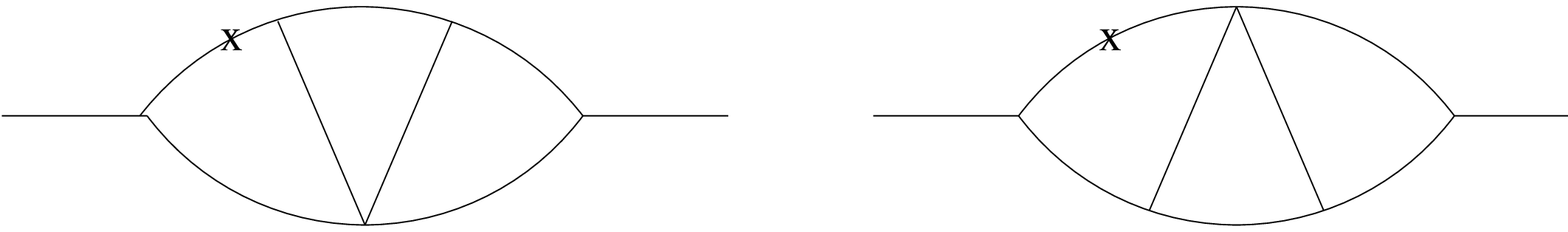}
\end{minipage}
\vskip -0.2 cm \phantom{W} \hskip 5 cm \textbf{Figure 14} \vskip
0.2 cm \noindent
Finally, topologies \textbf{B, C, D} are somewhat
nicer in that one can always get to the eight-propagator level by two
successive applications of the triangle rule. They contain the same master
integrals $M_{35}, \, M_{36}$ and only a \textbf{BU} topology with a
cross on the $p_4$ or $p_5$ propagator (c.f. Figure 3) comes in additionally.

The recursive procedure is quite laborious because a separate routine is
needed for the evaluation of every triangle solvable derived
topology with a numerator. Further, we had to create tables of
$M_{35}$ and $M_{36}$ with two dots and three dot products in the
numerator, likewise for the other non-elementary cases. These
$M_{35}, \, M_{36}$ integrals present a formidable computing
challenge because the matrix size is of the order $1650
* 4400$ in polynomials in $\epsilon$. It seems more appealing to
attempt a complete numerator reduction directly on the topologies
$\textbf{A, B, C, D}$ because there are no dots and one needs to
consider only four distinct mixed products in the numerator so
that the matrix results considerably smaller.

The advantage of our architecture is that the Gaussian elimination
need not be pushed to the end; it was usually sufficient to
eliminate less than half of the integrals in a basis to get
definite results for any given numerator term. We realised the
Gauss algorithm between \emph{ANSI C} and \emph{Mathematica}. The
whole problem addressed in this article could be solved on desktop machines
with moderate memory, although the total runtime amounted to several weeks.
A sample calculation with the programmes will be presented in a separate publication,
probably on the example of $M_{36}$.

We conclude this section with a nice observation on the
Laporta algorithm: The integrals \textbf{A, C} without any numerator
are of dimension $1/q^4$ (the fractional part is not important here) and thus
power-counting finite, and they remain so with numerator $p_i.q$. In
our programmes we had not substituted explicit results for the master integrals
till the very end, but rather kept the rational coefficient function that the
Gaussian elimination runs up in the corresponding columns. Finiteness of \textbf{A, C}
with any of the aforementioned numerators constrains the leading orders of the
master integrals. The \textbf{B, D} graphs without a numerator are
divergent due to their triangle subgraphs like e.g. the \textbf{BE} constellation at three
loops. With numerator $p_1^2$ or $p_2^2$ \textbf{D} becomes finite, on \textbf{B}
we may use any $p_i.q, \, i \in \{1 \ldots 4\}$.

The system of equations is strong enough to determine
\begin{eqnarray}
M_{35, \, 11111111} & = & \frac{1}{(4 \pi)^8} \biggl[ \,
\frac{\zeta(3)}{2 \, \epsilon^2} + \left(
\frac{\pi^4}{120} + \frac{11 \, \zeta(3)}{2} \right) \frac{1}{\epsilon} + \left(
- \frac{23 \, \zeta(5)}{2} + \frac{11 \, \pi^4}{120} + \frac{83 \, \zeta(3)}{2} \right)
\nonumber \\
&& + \left(\frac{59 \, \zeta(3)^2}{6} -\frac{2 \, \pi^6}{63} - \frac{253 \, \zeta(5)}{2}
+ \frac{83 \, \pi ^4}{120} + \frac{535 \, \zeta(3)}{2} \right) \epsilon + \ldots
\, \biggr] \biggl(\frac{q^2}{\tilde \mu^2}\biggr)^{- 4 \epsilon}  \, , \nonumber \\
M_{36, \, 11111111} & = & \frac{1}{(4 \pi)^8} \biggl[
\frac{5 \, \zeta(5)}{\epsilon} + \left( - 7 \, \zeta(3)^2 + \frac{5 \, \pi^6}{378}
+ 35 \, \zeta(5) \right) + \ldots \, \biggr] \biggl(\frac{q^2}{\tilde \mu^2}
\biggr)^{- 4 \epsilon} \, , \label{masters} \\
NO_{\epsilon1111111} & = & \frac{1}{(4 \pi)^6 \, q^2} \biggl[ \,
20 \, \zeta(5) + \left( 80 \, \zeta(3)^2 + \frac{10 \, \pi^6}{189} + 40 \, \zeta(5) \right)
\epsilon
+ \ldots \, \biggr]  \frac{ \left( q^2 \right)^{- 4 \epsilon}}{\left( \tilde \mu^2
\right)^{- 3 \epsilon} \left(\mu^2\right)^{- \epsilon}} \, .
\nonumber
\end{eqnarray}
in exact agreement with the literature \cite{masters} upon conversion to
the ``G-scheme''. The given orders of the $\epsilon$ expansion of the masters are also
all that our project required. The Laporta algorithm hence turned out to be
self-sufficient.

Finiteness constraints on the level of the nine-propagator integral in the
middle of Figure 12 determine one order less in $M_{35}, \, M_{36}$, and by finiteness
of $M_{35}$, $M_{36}$ constellations (we may introduce $p_i.q/p_i^4$ on any line) we can
only fix the first two terms in $M_{35}$ while no condition on $M_{36}$ ensues.

\section{Conclusions}

In this work we discussed the Konishi anomaly as an operator mixing problem in
${\cal N} = 4$ SYM with gauge group $SU(N)$. Up to normalisation the two operators
that mix are the unintegrated chiral superpotential $B$ and Yang-Mills action $F$ of
 the ${\cal N}= 1$ superfield formulation of the theory.
Both are not finite on their own, but the linear combination $O = F - 4 \, g \, B$
is protected. The second mixture is fixed by the conformal properties of two-point
functions, i.e. orthogonality to the protected operator and the form of its two-point
function. We considered the first two non-trivial orders in perturbation theory using
SSDR as a regulator and multiplicatively renormalised by Z-factors.
In the result
\begin{equation}
K \, = \, Z_B \, B + \frac{g \, N}{32 \, \pi^2} \, Z_F \, F
\end{equation}
both renormalisation factors acquire singular higher loop corrections. Further,
$Z_F$ also receives non-vanishing finite corrections at higher loops. Our work
fully confirms the general considerations about the singularities of the Z-factors put
forward in \cite{EJSS} and extends the leading order perturbative analysis presented
there. The descendant operator $K$ has the anomalous dimension of the Konishi
operator as required by supersymmetry.

In conclusion, in our framework the Konishi anomaly is not one-loop exact in
contradiction to the comment after equation (2.110) in \cite{67}.
As an explanation we remark that higher loop mixing coefficients are usually
scheme dependent. The approach quoted in \cite{67} supposes a definite
prescription for the renormalisation of the coupling constant appropriate to
general ${\cal N} = 1$ theories, while our scheme
is tailor-made for the conformal ${\cal N} = 4$ case. A general discussion
in rigorous perturbation theory wide enough to reconcile the two contrary
points of view is given in \cite{solution}.

The technically hardest part of the project was to elaborate the four-loop two-point
correlator $\langle \bar B F \rangle_{g^5}$, which we achieved by the Laporta
algorithm, i.e. integration by parts paired with Gaussian elimination. We hope to
separately publish the computer programmes developed to this end.
Apart from the interest inherent to the renormalisation properties of the Konishi
anomaly, the four-loop correlator is a vital piece of the calculation of the four-loop
anomalous dimension of the operator $K$ advocated in \cite{boxing} along the lines of
\cite{gamma3}; agreement with the existing result \cite{italovelizhanin} would at the
same time confirm the correctness of the method (thus the absence of a second ``anomaly''
in the supersymmetry variation of $K$) and further vindicate the aptitude of the
thermodynamic Bethe ansatz \cite{Glebetc} to describe ``wrapping corrections''
in the ${\cal N} = 4$ operator spectrum problem, i.e. the regime in which the dilatation
generator in the sense of \cite{dila} becomes inapplicable because the loop order exceeds
the spin chain length.

\section*{Acknowledgements}

The author is deeply indebted to K.~Chetyrkin for many discussions relating to the
Laporta technique and for sharing results on the master integrals prior to publication.
We are very grateful to A.~Pak for an independent test of the numerator reduction by
his own system. We thank S.~Moch for stipulating our interest in the approach
and K.~Sibold for comments on the manuscript. Much of the computer
work was done on a PC at the Spinoza Institute while the author was a member there.
M.~Segond pointed out \emph{jaxodraw} as a convenient way of drawing
Feynman graphs.

\section*{Appendix: Feynman rules and conventions}

\noindent The ${\cal N}=4$ SYM action formulated in terms of
${\cal N}=1$ superfields has the form
\begin{eqnarray}
  S_{{\cal N}=4}&=& \int d^4x \, d^2\theta \, d^2\bar\theta\;
\T \left(e^{gV} \bar \Phi_I e^{-gV} \Phi^I  \right) \label{action}  \\
  &+& \left[ \frac{g}{3!}\int d^4x_L \, d^2\theta\; \ep_{IJK} \T
(\P^I[\P^J,\P^K]) + c.c. \right] \nonumber \\
  &-&  \frac{1}{4 \, g^2} \int d^4x_L \, d^2\theta\; \T(W^\alpha W_\alpha)
\nonumber \\
  &-& \frac{1}{\alpha} \int d^4x \, d^2\theta \, d^2\bar\theta \,
\T \left( \Bigl(-\frac{1}{4} D^2 \, V \Bigr) \Bigl(-\frac{1}{4} \bar D^2 \,
V \Bigr) \right)
\nonumber \\
  &+&  \int d^4x \, d^2\theta \, d^2\bar\theta \, \T \left( (b+\bar b)
\, L_{\frac{g}{2} \, V} \, \Bigl( (c+\bar c) \, - \, \coth
L_{\frac{g}{2} \, V} \, (c - \bar c) \, \Bigr) \right) \, . \nonumber
\end{eqnarray}
The definition of the non-abelian field strength multiplet $W_\alpha$ is
\begin{equation}
W_\alpha \, = \, - \frac{i}{4} \bar D_{\dot \alpha} \bar D^{\dot \alpha} \left(
e^{g V} D_\alpha \, e^{- g V} \right) \, .
\end{equation}
We choose Fermi-Feynman gauge by putting $\alpha = 1$: the quadratic part of
the YM action becomes $+ 1/2 \int V \square V$. The action has the cubic
and quartic YM self-interaction vertices
\begin{eqnarray}
&+& \; \frac{g}{4} \, \int d^4x \, d^2\theta \, d^2\bar\theta \,
\T \left( \Bigl[ V, (D^\alpha V) \Bigr] \Bigl(- \frac{1}{4}
\bar D_{\dot \alpha} \bar D^{\dot\alpha} \, D_\alpha \, V \Bigr) \right) \\
&-& \frac{g^2}{12} \int d^4x \, d^2\theta \, d^2\bar\theta \,
\T \left( \Bigl[ V, (D^\alpha V) \Bigr]
\Bigl[ V, \Bigl(- \frac{1}{4} \bar D_{\dot \alpha} \bar D^{\dot\alpha} \,
D_\alpha \, V \Bigr) \Bigr] \right) \\
&+& \frac{g^2}{16} \int d^4x \, d^2\theta \, d^2\bar\theta \,
\T \left( \Bigl[ V, (D^\alpha V) \Bigr]
\Bigl(- \frac{1}{4} \bar D_{\dot \alpha} \bar D^{\dot\alpha} \,
\Bigl[ V, (D_\alpha \, V) \Bigr] \Bigr) \right) \, .
\end{eqnarray}
For completeness, the first two ghost vertices are
\begin{eqnarray}
&+& \; \frac{g}{2} \, \int d^4x \, d^2\theta \, d^2\bar\theta \,
\T \left( (b \, + \bar b) \, \Bigl[ V, (c \, + \, \bar c) \Bigr] \right) \\
& - & \frac{g^2}{12} \int d^4x \, d^2\theta \, d^2\bar\theta \,
\T \left( \Bigl[ (b \, + \, \bar b), \, V \Bigr] \, \Bigl[ V,
(c \, - \, \bar c) \Bigr] \right)
\end{eqnarray}
W.r.t. the definitions in \cite{bible}, the entire YM and ghost part of the
action changes sign, which is related to the spinor convention in the
integration measure. In addition, $V \leftrightarrow -V$.
The ghost propagators are
\begin{equation}
\langle \bar c(2) \, b(1) \rangle \, = \, \langle \bar b(2) \, c(1) \rangle
\, = \, \langle \bar \phi(2) \, \phi(1) \rangle \, .
\end{equation}

In Minkowski space the action (\ref{action}) appears in a weight
factor $e^{i S}$ under the path integral. Instead of putting the $i$ onto each
vertex we Wick rotate before determining the Feynman rules and
proceed in positive Euclidean signature.

The superfields carry an adjoint representation of the gauge group
$SU(N)$, and the (Hermitean) generators and the structure constants satisfy
the relations
\begin{equation}\label{69}
\T ( T^a T^b ) = \delta^{ab}\,, \qquad f^{abc}f^{abd} = 2 N \,
\delta^{cd}\, .
\end{equation}
The spinor convention is:
\begin{equation}
\psi^\a \, = \, \epsilon^{\a\b} \psi_\b, \qquad \psi_\a \, = \,
\epsilon_{\a\b} \psi^\b, \qquad \epsilon_{12} \, = \, 1, \qquad
\epsilon_{\a\b} \epsilon^{\b\gamma} \, = \, \delta_\a^\gamma \label{eq144}
\end{equation}
and exactly the same with dotted indices. Complex conjugation
replaces an undotted by a dotted index and vice versa; however, it
does not exchange lower and upper position.

The $2\times 2$ sigma matrices are Hermitian. The $\tilde \sigma$
matrix is obtained from $\sigma_{\a \dot \a}$ by raising of both
indices as defined by the last equation. They satisfy the
following relations:
\begin{eqnarray}
& \sigma^\mu \tilde \sigma^\nu \, = \, \eta^{\mu\nu} - i
\sigma^{\mu\nu} \, ,  &
  \tilde \sigma^\mu \sigma^\nu \, = \, \eta^{\mu\nu} - i \tilde \sigma^{\mu\nu} \, ,\\
& (\sigma^\mu)_{\a \dot \a} (\tilde \sigma_\mu)^{\dot \b \b} \, =
\, 2 \delta_\a^\b \delta_{\dot \a}^{\dot \b} \, ,  &
(\sigma^\mu)_{\a \dot \a} (\tilde \sigma_\nu)^{\dot \a \a} \, = \,
2 \delta^\mu_\nu \nonumber \, .
\end{eqnarray}
The partial spinor derivative obeys
\begin{equation}
\partial_\alpha \, \theta^\beta \, = \, \delta_\alpha^\beta \, , \qquad
\bar \partial_{\dot\alpha} \, \bar \theta^{\dot\beta} \, = \,
\delta_{\dot\alpha}^{\dot\beta} \, .
\end{equation}
Derivatives with upper indices are defined
by the raising rule given in the first equation in (\ref{eq144}) and the same with
dotted indices. (In case of doubt w.r.t. to signs the best strategy is always
to put the indices into standard position: up on spinors and down on derivatives.)
The superspace covariant derivatives are
\begin{equation}
D_\a \, = \, \partial_\a + i \bq^{\dot \a} \partial_{\a \dot \a},
\qquad \bar D_{\dot \a} \, = \, - \partial_{\dot \a} - i \q^\alpha
\partial_{\a \dot \a}, \qquad \partial_{\a\dot\a} = \partial_\mu
(\sigma^\mu)_{\a\dot\a}  \,.
\end{equation}
We define the square of a spinor without any weight factor:
\begin{equation}
\q^2 \, = \, \q^\a \q_\a, \qquad \bq^2 \, = \, \bq_{\dot \a}
\bq^{\dot \a},
\end{equation}
and similarly for the product of two different spinors.
Consequently,
\begin{equation}
D . D \; \q^2 \, \equiv \, - \frac{1}{4} D^2 \, \q^2 \, = \, 1,
\qquad \bar D . \bar D \; \bq^2 \, \equiv \, - \frac{1}{4} \bar
D^2 \, \bq^2 \, = \, 1.
\end{equation}
Under the $x$-integral we may thus identify
\begin{equation}
d^2\q \, = \, D . D, \qquad d^2\bq \, = \, \bar D . \bar D \, .
\end{equation}

SSDR means to modify the dimension of space, but to leave the
two-component spinor algebra untouched. In $D \, = \, 4 - 2 \,
\ep$ the basic propagator $\Pi_{ij}$ (\ref{piprop}) becomes
\begin{equation}
\Pi_{ij} \, = \, - \frac{\Gamma(1-\epsilon) \, \pi^\epsilon}{4 \pi^2
(x_{ij}^2)^{(1 - \epsilon)}} \, \delta(\theta_{ij}) \delta(\bar
\theta_{ij}) \, .
\end{equation}
The propagators carry Kronecker delta symbols for colour and flavour
indices, of course. We do not usually write these
in order to unclutter the notation.

\end{document}